\documentclass[manuscript,nonacm]{acmart}

\AtBeginDocument{%
  \providecommand\BibTeX{{%
    \normalfont B\kern-0.5em{\scshape i\kern-0.25em b}\kern-0.8em\TeX}}}
\usepackage{subfig}
\usepackage{float}
\usepackage{CJKutf8}
\setcopyright{acmcopyright}
\copyrightyear{2021}
\acmYear{2021}
\acmDOI{10.1145/1122445.1122456}

\acmJournal{PACMHCI}
\acmVolume{}
\acmNumber{}
\acmArticle{}
\acmMonth{}

\begin{document}

\title{End User Accounts of Dark Patterns as Felt Manipulation}


\author{Colin M. Gray}
\email{gray42@purdue.edu}
\author{Jingle Chen}
\email{chen2287@purdue.edu}
\author{Shruthi Sai Chivukula}
\email{cshruthi@purdue.edu}
\author{Liyang Qu}
\email{qu76@purdue.edu}

\affiliation{%
  \institution{Purdue University}
  \streetaddress{401 N Grant Street}
  \city{West Lafayette}
  \state{Indiana}
  \postcode{47907}
}

\renewcommand{\shortauthors}{Gray, et al.}

\begin{abstract}
  Manipulation defines many of our experiences as a consumer, including subtle nudges and overt advertising campaigns that seek to gain our attention and money. With the advent of digital services that can continuously optimize online experiences to favor stakeholder requirements, increasingly designers and developers make use of ``dark patterns''---forms of manipulation that prey on human psychology---to encourage certain behaviors and discourage others in ways that present unequal value to the end user. In this paper, we provide an account of end user perceptions of manipulation that builds on and extends notions of dark patterns. We report on the results of a survey of users conducted in English and Mandarin Chinese (n=169), including follow-up interviews from nine survey respondents. We used a card sorting method to support thematic analysis of responses from each cultural context, identifying both qualitatively-supported insights to describe end users' felt experiences of manipulative products, and a continuum of manipulation. We further support this analysis through a quantitative analysis of survey results and the presentation of vignettes from the interviews. We conclude with implications for future research, considerations for public policy, and guidance on how to further empower and give users autonomy in their experiences with digital services.
\end{abstract}



\maketitle

{\color{red} \textbf{Draft: October 20, 2020}}
\section{Introduction}
Digital products create and reinforce structures that guide our modern lives, providing a means to connect with others, share our experiences, perform mundane everyday tasks, and purchase goods and services. Modern accounts of these technologies range from the techno-optimistic to the downright dystopian---alternately channeling the power of technology to encourage equality and ease of access or the power of technology to corrupt and undermine individual agency. While modern Science and Technology Studies (STS) scholars point out that society has been shaped by the forces of marketing and advertising since the dawn of the industrial revolution, never before have systems been able to better track consumer behavior (via the ``surveillance economy'' \cite{Zuboff2019-kz}) and ``intelligently'' respond with customized and personalized prompts to optimize for monetary gain \cite{Mathur2019-ea,Luguri2019-bg,Susser2019-dm}. This desire for market and consumer control is not new, however; as far back as the 1960s, advertising designers recognized the tensions between contributing to societal good and working as shills for ``trivial purposes'':

\begin{quote}
    ``We do not advocate the abolition of high pressure consumer advertising: this is not feasible. Nor do we want to take any of the fun out of life. But we are proposing a reversal of priorities in favour of the more useful and more lasting forms of communication. We hope that our society will tire of gimmick merchants, status salesmen and hidden persuaders, and that the prior call on our skills will be for worthwhile purposes.'' \cite{Garland1964-cv}
\end{quote}

Of course, little has changed in modern advertising, even after this statement---known as the \textit{First Things First} manifesto---was renewed using similar language in 2000 \cite{firstthingsfirst2000}. Today, slick advertising appeals have been replaced by ``growth hacking'' approaches that seek to maximize business and shareholder value by acquiring clients at the lowest cost \cite{Bohnsack2019-py}. This market tendency---when paired with modern use of overtly persuasive and manipulative design patterns---has led to the emergence of ``dark UX'' and ``dark patterns'' as key drivers for building and sustaining a customer base \cite{Nodder2013-rw,Brignull2011-id,Brignull2013-ap,Brignull2015-qs,Gray2018-or}. These dark patterns are strongly linked to their manipulative counterparts in advertising over the past 50 years, while taking form in increasingly rapid and tailored ways \cite{Narayanan2020-va}.

In this paper, we build upon the concepts of manipulation and persuasion from the perspective of the technology end-user, seeking to supplement accounts of design ethics from designer \cite{Nodder2013-rw,Bowles2018-gx}, organizational \cite{Gray2019-ep,Shilton2013-dq,Shilton2018-sw,Steen2015-qw}, educational \cite{Coldewey2018-dx,Wong2017-wt,Wong2019-rc}, and professional ethics \cite{Buwert2018-uw,Gotterbarn2018-ja} perspectives with a description of how these persuasive systems impact user experiences and point towards felt qualities of manipulation. We seek to connect the discourses of dark patterns with end user experiences, with the goal of better articulating potential uptakes for public policy, user empowerment, and design ethics.

We report on the results of a survey study distributed to English and Mandarin Chinese-speaking participants (n=169), augmented by nine follow-up interviews. Through these data collection methods, we sought to elaborate participants' experiences of digital technologies that they felt were manipulative, including the qualities that caused them to suspect manipulation at varying levels of awareness, the emotions they felt when encountering these manipulative digital products, and the ways in which these participants were aware of the creation and creators of these manipulative experiences. While we did not directly ask participants about their experiences with ``dark patterns,'' we used the participants' ``felt manipulation'' as a proxy to investigate how dark-patterns-informed digital products might be experienced by end users. Building upon our survey and interview results, we offer an end user perspective on dark patterns, building on minimal existing literature from Maier and Harr~\cite{Maier2020-yk} in this space to identify opportunities to better support user agency through public policy and designer/organizational practices.

The contribution of this paper is two-fold: 1) We document the felt experiences of manipulation shared by users of digital products, providing a new perspective on ``dark patterns'' that facilitates further work that builds upon the intersections of user trust, designer intent, and public policy. 
2) We describe users' perceptions of the creator of these manipulative interfaces or technology systems from multiple perspectives, including the projected role of designer ethics and responsibility in the creation of manipulative technologies.


\section{Background Work}

\subsection{Manipulation in Digital Systems}

Manipulation has been studied in numerous disciplinary contexts, often building upon and further specifying the conditions by which an act or experience is manipulative---beyond what we already colloquially understand to ``feel'' or ``be'' manipulative. From a philosophy perspective, Ackerman~\cite{ackerman1995concept} contends that for an act to be considered manipulative it must first involve ``getting someone to go against what he finds natural or appropriate or is otherwise inclined to do.'' From a critical discourse perspective, van Dijk~\cite{van2006discourse} further positions manipulation as simultaneously dealing with ``a form of social power abuse, cognitive mind control and discursive interaction''---implicating both social dimensions of interaction and a bi-directional act of discursive practice that is shaped ``by cognitive and social dimensions.'' From a legal perspective, Waldman~\cite{Waldman2020-km} focuses on aspects of cognitive bias that impact the kinds of strategies that digital products use to manipulate user behaviors. Waldman \cite{Waldman2020-km} links these legal concerns to their description of common psychological forms of manipulation that impact users' decision making processes, including: anchoring (``the disproportionate reliance on the information first available when we make decisions''); framing (``the way in which an opportunity is presented to consumers---namely, either as a  good thing or a bad thing''); hyberbolic discounting (``the tendency to overweight the immediate consequences of a decision and to underweight those that will occur in the future''); overchoice (``the problem of having too many choices, which can overwhelm and paralyze consumers''); and metacognitive processes (e.g., ``perceiv[ing] difficulty as a signal of importance''). 

In an HCI and CSCW context, relatively little work has \textit{directly} addressed manipulation as a construct in relation to digital systems, although such a concept is clearly implicated in the study of restrictive online community practices, disinformation campaigns, and even in mechanisms used to discourage direct action against platforms. Frequently, ``nudging'' is used to describe a light form of manipulation whereby a system maintains user agency but privileges one type of behavior or outcome over another~\cite{wilkinson2013nudging,Weinmann2016-oo}. As stated by Weinmann et al.~\cite{Weinmann2016-oo}, ``Even simple modifications of the choice environment in which options are presented can influence people’s choices and `nudge' them into behaving in particular ways,'' concluding that ``there is no neutral way to present choices.'' In contrast, other scholars such as Fogg \cite{Fogg2003-rg} have actively called for the identification and harnessing of principles of persuasion, with the contention that these principles can be used to promote user experiences where behavior modification is desirable (e.g., increasing motivation, reducing addictive behaviors). However, the liminal spaces among manipulation and persuasion are actively in contention, with arguments over the role of user agency, social good, and digital products. Recently, Susser et al.~\cite{Susser2019-dm} have described the intersection of many of these concerns in relation to manipulation on online platforms, identifying the potential harm and loss of autonomy that these experiences foster. In this paper, we particularly rely upon Susser et al.'s \cite{Susser2019-dm} definition of manipulation: ``manipulation is hidden influence. Or more fully, manipulating someone means intentionally and covertly influencing their decision-making, by targeting and exploiting their decision-making vulnerabilities.'' While in the original source, these authors also seek to distinguish manipulation from persuasion, nudging, and coercion, due to the lack of literature relating user experiences of felt manipulation to existing known dark patterns, we use the term \textit{manipulation} in a broader sense, while also recognizing the need for more work in this area to connect end user experiences with designer intent. We also rely upon recent work that describes the role of crafty psychological nudging through the incorporation of ``dark patterns'' \cite{Brignull2011-id,Mathur2019-ea,Gray2018-or} and instances where coercion is blatantly described in ways that often force users to make decisions against their own best interest---what Gray and colleagues describe as ``asshole designs,'' \cite{Gray2020-zq} building on a subreddit by the same name.


\subsection{Dark Patterns and the Optimization of Shareholder Value}

A surge of work in the past five years has focused attention on persuasive, manipulative, and coercive practices in the design of digital products, with the moniker of ``dark patterns'' appearing as the most frequently occurring and compelling label for practitioner and legal discourse around technology ethics. The neologism of \textit{dark patterns} was coined in 2010 by UX practitioner Harry Brignull, who also holds a PhD in cognitive science. This concept brought together interest in persuasive technologies (e.g., captology \cite{Fogg2003-rg}), the rise of user experience (UX) and user research as means of better understanding capabilities and needs of users, and the rise in optimization and personalization of systems. Gray et al. \cite{Gray2018-or} summarized and extended a typology of dark patterns strategies in 2018, expanding upon Brignull's initial definition to identify these patterns as instances where ``user value is supplanted in favor of shareholder value.'' While Brignull's initial typology of dark patterns included mostly context-specific instances \cite{Brignull2015-qs}, largely drawing on examples from e-commerce, the work of Gray et al. \cite{Gray2018-or} shifted this language to derive five strategies that designers can take in order to create manipulative experiences that foreground shareholder value: nagging, obstruction, sneaking, interface interference, and forced action. Gray et al. \cite{Gray2020-zq} have recently described connections between dark patterns and ``asshole design,'' identifying the former strategy as a sneaky and surreptitious form of persuasion, while the latter properties reveals themselves in overtly coercive ways.

Building on both Brignull and Gray et al., a wide range of work has been published in the last three years by academics and practitioners alike, drawing additional attention on the misuse of knowledge about users to coerce, manipulate, or persuade (e.g., \cite{Narayanan2020-va,Greenberg2014-dg}). In the design practitioner space, including conversations about design intent and responsibility \cite{Tollady2016-pa,Wong2020-ja} and identification of specific contexts where dark patterns have been employed \cite{Vance2016-yv,Schlosser2015-o,Read2018-cr}. In the academic space, scholars have addressed dark patterns in gaming \cite{Deterding2020-fx,Zagal2013-ms}, robotics \cite{Lacey2019-cy}, and proxemic sensing \cite{Greenberg2014-dg}. Collections of anti-patterns \cite{MacDonald2019-nq,Mirnig2017-ij}, privacy patterns \cite{Fritsch2017-re}, and bright patterns \cite{Grassl2020-lh} have continued to expand the dark patterns literature. A newer line of scholarship addresses the incidence of dark patterns in relation to data privacy and security, including investigation of legal issues relating to consent \cite{Soe2020-se,Nouwens2020-ij,Grassl2020-lh,Luguri2019-bg,Gray2020-zf}, dark patterns in ecommerce \cite{Mathur2019-hx,Moser2020-sf}, and manipulation in mobile applications \cite{Di_Geronimo2020-mh}. Increasingly the notion of dark patterns is being used to connect scholarship across legal, interaction design, user experience, privacy and data protection, and ethics domains---foregrounding the need for users to have agency as they use digital systems. 

These efforts in describing how dark patterns can result in a lack of transparency or user agency has been raised as a space for new policies to be formed. In the European Union, legal action in response to the General Data Protection Regulation (GDPR; \cite{noauthor_undated-gy}) has been leveraged to indicate where dark patterns may limit free and unambiguous consent \cite{Soe2020-se,Gray2020-zf}, while similar data protection measures in the United States (e.g., the California Consumer Privacy Act \cite{noauthor_2018-ww}) may yield a similar platform for increased user protection, alongside experimental work by legal scholars (e.g., \cite{Luguri2019-bg}). The DETOUR act has also been proposed in the United States Senate as a means of banning certain forms of dark patterns \cite{Kelly2019-me}.

Of particular note in relation to this current paper is the work of Maier and Harr \cite{Maier2020-yk}, who conducted a set of interviews and focus groups in order to describe users' experiences and perceptions of dark patterns. This paper concluded that perception related to user impressions, assessment, balance, and acceptability, pointing towards ongoing sensemaking that users engage in to determine if or to what degree they are being manipulated. We explicitly build upon this work, using a survey study with English-speaking and Mandarin Chinese-speaking users to reveal broader patterns and examples of users' perception of manipulation.

\section{Our Approach}
We used a mixed methods approach to describe end users' experiences and perception of manipulative technology interfaces, collecting user responses through a survey study and a follow-up interview study with interested survey respondents. We collected data from English-speaking and Mandarin-speaking users, not primarily as a means of ``proving'' cross-cultural differences, but rather as a way to problematize the notion of manipulation across cultural and geographic boundaries. It is not our ambition to directly compare data arising from these unique contexts, but rather to identify and articulate as broad a set of interpretations of manipulative user interface practices as possible. Through these efforts to describe the felt experiences of users in relation to their daily interactions with technologies, we answer the following research questions:

\begin{enumerate}
\item What makes digital products feel ``manipulative'' to end users?
\item What emotions do end users relate to past manipulative experiences with digital products?
\item How aware are the users about the creation and creators of manipulative technologies?
\end{enumerate}

In this section, we detail our research contexts, data collection methods for the survey and interview studies, and mixed methods data analysis. 

\subsection{Survey Design}
We conducted a survey study to identify end users' daily interaction with manipulative interfaces, emotional reactions to the manipulation, and perception about creators of these technology manipulations. We deployed a 22 question survey that included a mix of multiple choice questions and open ended questions to allow participants to illustrate their \textit{felt} experiences of manipulation. The sections of the survey aimed to record: agreement to participate in our study, demographic information of the respondent (gender identity, age, profession, country of residence), patterns of daily usage of smartphone application or website on their personal digital devices, open-ended examples of manipulative technology interactions (e.g., \textit{``What makes you feel mistrustful of a smartphone application?''}), their immediate and lasting emotional reaction to this instance of manipulation (listing negative emotions derived from psychology studies on emotion~\cite{Cowen2017-fn,feelingwheel} and negative affect~\cite{Watson1988-po}), their perception of being valued as a person or customer (drawn from a study on dark patterns \cite{Gray2018-or}), and their awareness of the creators of manipulative interfaces.  
We concluded the survey asking respondents if they would be interested in participating in a follow-up interview. 

As a part of validating our survey protocol, we conducted a pilot study with over 50 participants to receive feedback about the average time taken to complete the survey and the framing of the questions. The survey protocol was then modified based on feedback from the pilot responses and validated in English on Qualtrics by a group of researchers that represented various forms of practitioner and design engagement. After finalizing the protocol in English, the protocol was translated into Mandarin Chinese by a researcher whose native language is Mandarin. Then the translation was reviewed by another native Mandarin-speaking researcher to further improve its accuracy. 

\subsection{Survey Distribution}
The final survey was distributed through the personal networks of the research team and through social media, including WeChat and Facebook groups about shopping and parenting. Our goal was to target ``everyday users'' who regularly used digital products but did not have detailed knowledge about the design or development of these products. By finishing the survey, participants had a 1 in 50 chance to win a \$10 USD online shopping gift card. For Chinese-speaking participants, we provided them with a 1 in 50 chance to win WeChat currency of the same value (around 70 RMB). After data collection was complete, we downloaded the responses into a spreadsheet for further analysis. 



\subsection{Summary of Survey Responses}
In total, we received 253 responses to the survey, including 169 respondents that completed the survey fully (a 65.61\% complete response rate) and met our inclusion criteria. Out of 169 completed responses, the distribution of participants geographically was as follows: China (n=103), United States (n=40), United Kingdom (n=9), Canada (n=5), and other (n=13).  As our study was not particularly focused on comparative analysis based on geography, we included all English-speaking responses together for further analysis. Gender identity distribution of participants was as follows: female (n=108; 61.7\%), male (n=62; 35.6\%), and did not disclose (n=4; 2.3\%). Age distribution of participants in our data set was as follows: 18-24 yrs (n=19), 25-34 yrs (n=39), 35-44 yrs (n=19), 45-54 yrs (n=58), 55-64 yrs (n= 30), 65-74 yrs (n=2), and did not disclose (n=7). Removing the 7 responses that did not disclose their age, the average age of the participants is 42 years, with a standard deviation of 12.89 years. 
The survey participants had substantial familiarity with technology; when reporting on the combination of mobile and desktop device use, 82.2\% (n=143) of the participants identified themselves as heavy technology users (between 5-14 hours of use per day) and only 15\% (n=26) as lighter technology users (1-3 hours of use per day). The survey respondents were from a range of professions and backgrounds including freelancers, homemakers, students, printers, nurses, nutritionists, money lenders, engineers, software trainers, technicians, attorneys, administrative assistants, therapists, business executives, and CEOs. 




\subsection{Follow-up Interviews}
 After collecting our survey responses, we conducted a follow-up interview with a subset of American and Chinese participants. We identified this subset by balancing demographic characteristics and experiences with technology in each cultural context, resulting in five Mandarin speakers and four English speakers
 . Each interview was approximately 15 to 30 minutes in duration, and was conducted by one or two researchers. All interviews were conducted in the native language of the participant. We used a critical interview approach~\cite{Carspecken1996-bq}, seeking to guide participants to share their point of view and subjective experiences with technology. The first section of the interview was framed by the participant's survey responses, including more detail on their stories of being manipulated, and what aspect(s) of technology felt manipulative or unduly persuasive to them. In the second section, we focused on the participant's perception of the designer of manipulative technologies, asking questions such as: ``Why do you think someone created this manipulative experience?'' and ``How do you think the designer knows how to persuade you to act in a certain way?'' Participants were encouraged to provide stories and specific experiences to support their answers, and we asked follow-up questions to facilitate the conversation. The interviews were voice-recorded with the permission of the participant and were transcribed for analysis. We used online transcription tools for English speaking participants and manually transcribed recordings of Mandarin speaking participants. We retained the dialogue in Mandarin and provided excerpts in the same dialect to retain the authenticity of the communication and emotions behind user stories. 










\subsection{Data Analysis}
Given the different data collection methods, we adopted a mixed-methods analysis approach including: 1) card sorting to analyze survey responses for open-ended questions, answering RQ\#1; 2) quantitative analysis to analyze choice or scale-based questions, answering RQ\#2 \& RQ\#3; and 3) case analysis to present vignettes from interview narratives, elaborating RQ\#3. We detail each of these analysis procedures in the subsections below. 


\subsubsection{Card Sorting}
The open-ended questions in the survey prompted a range of examples and stories by the participants in both Chinese and English. Having received responses from both multiple contexts and languages, we  used card sorting as a means of foregrounding cultural complexity in relation to manipulative technologies. To prepare for the physical card sort activity, we separated 169 survey responses into 609 data units. Each response was printed on a 5 inch square piece of paper, which we used as `cards.' The cards were not separated by language; rather, we mixed them together and treated them equally. This activity was conducted by four researchers: one from the USA, two from China, and one from India. All researchers were trained in qualitative thematic analysis and card sorting methods through previous design and research projects. We conducted a physical card sort as a stand-up activity on a large table, with sufficient room to move around and display as many cards as possible. We collaboratively created categories as each researcher read off their cards, translating if necessary, and discussing our collective interpretation. We also occasionally made use of the Google Translate app to augment the Chinese speakers' interpretations and further discuss the preliminary themes. After we sorted approximately 100 cards, all members began to have a shared understanding of the categories and felt the categories had been exhausted.  The physical activity of sorting cards allowed us to replicate the cards wherever required and place them under multiple groups. The physical sorting also allowed us to position the responses in the form of a continuum to represent a sense of progression of manipulation based on the examples shared. We present this spectrum in Section~\ref{sec:continuum}. The two Mandarin speakers then focused mainly on Chinese responses, while the other researchers categorized most of the English responses and assisted with the grouping of Chinese responses based on Chinese characters that had become familiar due to their frequency. Over a three and a half hour session, this initial round of cross-cultural card sorting allowed us to condense the 609 data sets into 12 preliminary themes, which were later condensed into salient themes based on the ``felt '' manipulation of the users' experiences with digital products to answer RQ \#1 in Section~\ref{sec:themes}.

\subsubsection{Quantitative Analysis}
We conducted basic descriptive statistical analysis on the choice or scale-based questions asked in the Survey. We did the following types of statistical analysis to answer RQ\#2 \& RQ\#3: 1) To answer RQ\#2 in Section~\ref{emotion}, we analyzed the frequency of responses to the listed emotions across the scale of being affected ``Very Slightly or Not at all'' to ``Extremely.'' We also plotted the most frequently felt emotions by calculating frequencies of ``Quite a Bit'' and ``Extremely'' against different age segments; 2) To answer RQ\#3 in Section~\ref{sec:awareness}, we analyzed and plotted the frequencies of users' awareness of being manipulated across each age segment, including who they blamed for the manipulation; and 3) To answer RQ\#3 in Section~\ref{Sec:creators}, we analyzed and plotted the frequencies of how users felt they were valued in general and without awareness of being manipulated. In the respective sections, we have provided further detail on our calculations.

\subsubsection{Case Analysis}
We collected vignettes of users' narratives as they described the creators of manipulation shared through semi-structured interviews. To elaborate further on the participant's level of awareness (Section~\ref{sec:awareness}) and their perceptions of the creators of manipulative products Section~\ref{Sec:creators}, we present three vignettes to describe each end user's felt experience of manipulation, who they blamed, and the ways in which they wished to respond to the creators of that manipulation. These cases are not meant to be conclusive and representative of the whole data set, but rather as a means of raising examples and reactions to manipulative experiences to frame our findings and point to opportunities for future work. 
 
\section{RQ\#1: Describing End Users' Felt Experiences of Manipulation}
To evaluate the users' experiences of manipulation in relation to digital products, we first identify what aspects of technological systems may help the user to identify that they are being manipulated (Section~\ref{sec:themes}. We then build upon this analysis with a continuum of manipulation with a progression of characteristics of manipulation sorted along a temporal progression, from initial impressions to longer-term user interaction (Section~\ref{sec:continuum}).

\subsection{Perceptions of `Manipulation'}
\label{sec:themes}
First, we seek to describe what aspects of technological systems feel ``manipulative'' to end users. In the following sections, we describe a range of areas that impacted this felt manipulation, including: distrusting the digital product (n=103), feeling concern about the privacy of their personal information (n=72), knowledge of being tracked without explicit notification as part of a larger ``big data'' threat (n=51), identifying barriers to feeling secure (n=45), being aware of explicit manipulation tactics (n=28), and the use of freemium products (n=21). 

\subsubsection{Overall Perception of Trust or Distrust} 
In this theme, manipulation is perceived through the lens of who or what users trust and distrust (n=103). Lacking trust in the digital product can lead users to feel as if they are being manipulated, even if the specific location of this feeling of trust or distrust is less clear. Users from both cultural contexts mentioned a sense of insecurity when interacting with ``unfamiliar'' digital technologies. In these cases, some users tended to pay more attention to the particular characteristics of manipulative digital technologies (see Section~\ref{sec:continuum}) that these users may have identified based on their previous experiences, leveraging this knowledge to actively find evidence to prove their assertion that this product is created by ``scammers'' or \begin{CJK*}{UTF8}{gbsn}``骗子 (swindlers)''\end{CJK*}. However, if the users felt that the digital product sought to build a sense of trust, users were more likely to use information from these product(s) as a reliable information source. In fact, it appears that both Mandarin and English speaking users frequently relied on external information sources to identify the potential of manipulation, using this judgment to impact their overall trust of the digital product. 
If one or more of these sources suggested that a digital product was manipulative or insecure, the users appeared much more willing to label the digital product as ``manipulative software.'' Users may certainly trust one source more than the other, but 
We identified some common information sources that generally appeared in both cultural contexts, including: friends, public broadcasting, reliable software (e.g. browser, anti-virus alerts), and user ratings/reviews from certain trusted platforms. Interestingly, we discovered that a cultural bias may affect the perception of who/what are unreliable; English-speaking participants often used peripheral cues (e.g., ``website that is hosted in a different country'' and ``poor developer documentation'') as a way to avoid potentially manipulative products. In parallel, Chinese participants were unlikely to trust \begin{CJK*}{UTF8}{gbsn}`` bundled software (捆绑软件)''\end{CJK*}, which refers to instances when related yet unwanted software was installed alongside the desired application on the user's personal device. 


\subsubsection{Collection of Personal Information}
In this theme, users perceive manipulation in relation to a lack of privacy as they interact with digital products (n=72). When a digital product demands or collects a user's personal information in an aggressive way, users may feel that the product is challenging their privacy in a manipulative way, complicated by instances where users cannot even comprehend ``why [the digital product] need[s] my info.'' 
Responses from users in this theme revealed that users are not always resistant to providing basic information (e.g., name, email address) to a digital product in exchange for convenience or value. However, users expressed the existence of a threshold when a digital product tries to gather ``too much or unnecessary'' personal information. Often, this threshold was reached when digital products seek to collect information that could be deemed sensitive. Sensitive information mentioned by participants included: a mailing address, contact information, \begin{CJK*}{UTF8}{gbsn}``身份证号码  (a personal identification number) or Social Security Number'',\end{CJK*} and payment-related information. 
Similarly, asking for an unreasonable level of permissions on an app was also seen as a way that a digital product could demanding personally information ``aggressively.'' Users were especially skeptical and sensitive when digital applications asked for location and contact list permissions, which tend to disclose more private information about users. Culture also plays a potential role in this theme, which provided us interesting insights about how digital product manipulate users to provide personal information. Some Mandarin speaking participants mentioned a concern about  \begin{CJK*}{UTF8}{gbsn}"钓鱼网站 (phishing websites)"\end{CJK*}, which often brand themselves to look similar to another trustworthy website and lead users to input sensitive information. English speaking participants also mentioned being wary of a ``long user agreement'' in which companies often hide unfair use and collection of a user's information.


\subsubsection{Threats of Big Data}
Building on the previous theme, users' concerns about the potential threats of Big Data in relation to privacy concerns and data collection were also a common way that users perceived a manipulative threat (n=51). 
Instead of emphasizing the collection of information that users deemed ``sensitive,'' this theme focuses on how users perceive the tracking of their interactions with digital products and how product stakeholders may use these interaction data to achieve business goal without explicitly notifying users in ways that felt manipulative. Although users in both cultural contexts were aware that ``data theft'' exists and that digital products are \begin{CJK*}{UTF8}{gbsn}"窃听后台分析数据 (tracking and analyzing the data in the back end)"\end{CJK*}, participants didn't characterize their digital footprint in detail or describe how it could be tracked on an abstract level. Instead, our participants provided details about how these interaction data were concretely utilized by products and services. Two main types of uses that felt manipulative were identified: advertisements and search results. Some users mentioned having been shown ``ultra-specific advertisement'' based on their interests. The display of these advertisements could be based on users' interaction within the same digital product that displayed the advertisement. However, based on users' impressions, the knowledge of user's interest usually came from other digital products, which supported some users' claims that their ``data was sold for business purposes.'' Search results were identified to be another common usage of user's data. For example, results may be displayed based user's interests rather than the keyword(s) that the user input, which cause ``inaccurate results of search.'' Additionally, a Mandarin speaking participants indicated that \begin{CJK*}{UTF8}{gbsn}"搜索结果全部指向百家号 (the search results are all related to BaiJiaHao, a media platform owned by the largest Chinese search engine Baidu.)''\end{CJK*}


\subsubsection{Barriers to get security}
When users felt their property or information was directly threatened because of using a digital product, users would feel they have been manipulated. We categorized this theme as ``Barriers to get security'' (n = 45). There were three major security barriers that were identified by participants from both cultural contexts: they are fraud or scam, virus, and hacking. Unfortunately, our respondents didn't provide us further details for each of these sub themes. However, what we discovered was Mandarin speaking participants brought up more concerns about these security related issues, such as \begin{CJK*}{UTF8}{gbsn}"不良软件 (illegal software)" and “黑网站 (fake/unreliable website)”\end{CJK*}.


\subsubsection{Awareness of Explicit Manipulation}
In this theme, manipulation was identified in relation to an explicit and specific form of user manipulation that participants had experienced (n=28). A small group of participants explicitly listed a specific form of user manipulation \begin{CJK*}{UTF8}{gbsn}``套路 (tricks)''\end{CJK*} or user knowledge that could be used to create manipulative experiences. Our survey responses indicate that several users appeared to be aware of some common manipulation techniques and mechanisms, and when they encountered a similar situation, they could quickly identify these potential forms of explicit manipulation. As one example, a respondent discovered that ``the placement of the agree button is often where the enter to continue button is (placed),'' reflecting location-based dark patterns that are commonly deployed by developers. Designers rely upon manipulative design techniques such as these to trick, manipulate, or even coerce users into accepting rules or policies that most users may not agree in other cases, since many users often have the tendency to click buttons such as ``Next Step'' or ``Continue'' without reading the content carefully. 


\subsubsection{Use of Freemium Products}
In this theme, we identified that some users were especially sensitive to freemium products (n=21), which include digital products or services that require user payment to proceed or maintain access after using the product or service for free. Requirement of payment or \begin{CJK*}{UTF8}{gbsn}"收费 (require to pay)"\end{CJK*} often caused users to view a digital product as potentially manipulative, and even the ``possibility of in-game (or in-app) purchase'' could result in questioning the trustworthiness of a digital product. This high sensitivity appeared to be related to the lack of direct and precise disclosure of the digital product's business goal to users, which has been questioned by users in both cultural contexts. Freemium products were reported to be carefully crafted by its creator, with the potential for disruption of the user's experience used as a strategy to manipulate users to pay. This appeared in the form of ``payment traps,'' ``getting me to pay to unlock digital items,'' \begin{CJK*}{UTF8}{gbsn}``使用过程中强制性收费 (mandatory fees in the middle of usage),'' and ``以各种目的让你付钱 (using every excuse to let you pay).''\end{CJK*} Users also felt that this manipulation could be more intensely felt when there was a mismatch between user's expectation and reality: for example, instances where the freemium product was advertised as free to use.


\subsection{Temporal Progression of ``Felt'' Manipulation}
\label{sec:continuum}
Through discussion, we discovered that visual and interactive characteristics could contribute greatly to end users' determination of technologies' trustworthiness. To investigate this behavior further, we organized the instances along a continuum from initial and general impressions of a product to extended use and interaction with the product (Table~\ref{continuum}). We report on each element of the continuum in greater detail in the following subsections.

\begin{table*}[h]
\centering
\begin{tabular}{ p{14em} p{32em}  }

\toprule
\textbf{Characteristic} & \textbf{Definition} \\ 
\midrule
\textbf{Initial Judgment (n=24)} & Initial negative impression or judgment. \vspace{5px} \\
\textbf{Inspection of Details (n=82)} & Negative impression after inspection of interface details. \vspace{5px} \\
\textbf{Felt Persuasion (n=34)} &  Negative impression after felt technology persuasion.  \vspace{5px} \\
\textbf{General Conclusion (n=14)} & Conclusion of negative sentiment after considered interpretation of usability and usefulness. \vspace{5px} \\
\textbf{Undesired Interaction (n=30)} &  Conclusion of negative sentiment after undesired or unnecessary interactions occur. \vspace{5px} \\
\textbf{Negative Results from \newline Interaction (n=24)} &  Conclusion of negative sentiment after user interaction results in unwanted outcomes. \vspace{5px} \\
\bottomrule
\end{tabular}
\caption{Specific characteristics of manipulation (n=208): the top of the table is focused on initial and general impressions, moving towards the bottom which is focused on direct user interactions.}
\label{continuum}
\end{table*}

\subsubsection{Initial Judgment}
General impressions about a technology often relied upon immediate  \begin{CJK*}{UTF8}{gbsn}``直觉 (intuition)'' that the digital product was manipulative, supported by a feeling that ``something seemed to be wrong'' or judgments that the digital interface ``looked unprofessional (看起来不正规)''\end{CJK*}. Some English-speaking participants augmented this understanding by mentioning ``bad or flashy design'' as a potential source of early judgment that a digital product might be manipulative. 

\subsubsection{Inspection of Details}
Specific details about how the content was displayed was often the focus of users after their initial judgment. Advertising was a dominant focus of attention by users, and based on our analysis, end users in both cultural contexts did not appreciate ``advertising driven'' products, and the presence of advertising brought additional scrutiny. There were three aspect of ads that were brought up by our participants to identify ``advertising driven'' interfaces: quantity (e.g., ``too many advertisements''), quality (e.g., ``sketchy advertisements''), and format (e.g., ``intrusive ads'' or ``pop ups''). Beyond advertisements, other details of a user interface could also raise a red flag for users, such as ``whenever it has the option to link to Facebook, Google etc.,'' \begin{CJK*}{UTF8}{gbsn}``出现不健康信息和图片'' (the appearance of contextually inappropriate content)\end{CJK*}, and ``misspellings.'' One participant who specifically focused on the more technical aspect of the technology identified that they ``read the code'' or ``monitor[ed] the traffic'' as a way of inspecting for manipulation.

\subsubsection{Felt Persuasion}
Experiencing and identifying direct persuasion was another determinant for users to not trust a technology. 
Many people identified that they were extremely careful when they detected ``extreme statements'' or content with strong biases, such as ``news articles that are mainly used to support political opinions,'' ``obviously twisted and misleading information,'' and other similar instances. These forms of content were seen to be ``clearly playing on emotions and political leanings'' a person may have. ``Stories that seem too good to be true'' also raised concerns among our participants because people felt that these statements were disingenuous. Another type that was identified was offering incentives for doing something or even nothing. Some websites encouraged users to share their information by promising ``free'' goods or financial benefits, including ``[a] website [that] wanted me to sign up in order to see the content.'' Some sneakier methods of persuasion that are commonly used were also identified by some users; for example, \begin{CJK*}{UTF8}{gbsn}``闭合性的选择''\end{CJK*} represents a situation where the digital product provided users with limited options which are similar or related to each others. No matter which one the user chose, it would always benefit the creators of that digital product.

\subsubsection{General Conclusion}
People were also able to evaluate the technology after initial interaction, using their initial judgment alongside this interaction to come to a reasoned conclusion. Instances such as \begin{CJK*}{UTF8}{gbsn}``无用 (no direct usage)''\end{CJK*} and sentiments that the site was ``not user friendly'' demonstrate the capacity of users to reach a generalized conclusion about the site. Some users also concluded that digital products were not useful to them or had a bad user experience, linking behaviors such as \begin{CJK*}{UTF8}{gbsn}``使用麻烦 (complicated to use)''\end{CJK*} to the potential for manipulation.

\subsubsection{Undesired Interactions}
After more in-depth engagement with the digital product, users tended to evaluate the desirability of certain interactions in relation to the potential for manipulation. When the experience was ideal, users were able to locate the content they expected to see and could predictably interact with the product. However, if ``something undesired happened'' when the user was trying to accomplish a specific goal, the user may feel a sense of manipulation. Many participants identified more specific causes for these undesired interactions, which were often designed intentionally to produce unwanted results for users. For instance, one user mentioned ``click[ing] on [an] article (and the article) turns into ads", while other users mentioned \begin{CJK*}{UTF8}{gbsn}``不经意跳转各种奇怪的网页 (inadvertently jumping to various weird web pages),'' ``强制跳转其他网站 (forcing me to go to other websites)''\end{CJK*}, and ``interrupting ads to buy more things'' as undesirable behaviors that resulted in an assessment that the product was being manipulative. 
Beyond these intentional design choices that pointed towards manipulation, some users stated system errors, such as ``bugs'' or ``glitches,'' were another cause for some of their undesired interactions, and could lead to an assessment of quality that intersected with manipulative intention.

\subsubsection{Negative Results from Interaction}
Some users only realized that they were manipulated after interacting with the technology for a period of time, after experiencing negative consequences or being harmed by interacting with a manipulative user interface. Our participants shared a wide variety of these negative results, including ``using lots of data,''  \begin{CJK*}{UTF8}{gbsn}自动下载  (downloading [unwanted software] automatically), ``影响系统运行 (affecting operation system running),'' and ``影响工作与学习 (affecting my work and life).'' The ``roach motel'' was a common pattern that resulted in delayed and negative results; ,any participants listed situations such as ``无法轻松删除 (unable to delete easily),'' ``无法退出 (unable to exit),'' ``opt out pathways were very difficult to find,'' and ``subscription trap'' as examples of manipulation that may take extended use to discover. \end{CJK*}

Finally, some users came to the conclusion that they were being manipulated only after they experienced negative impacts from \textit{extended} interactions with a digital product. One common strategy that emerged in this extended use context was gamification, which was used to ``encourage'' or ``nudge'' users to perform undesired tasks repetitively for certain benefits---combining 
the notion of \begin{CJK*}{UTF8}{gbsn}``performing tasks to gain something (完成任务获得一些东西)''\end{CJK*} and other culturally-specific examples such as 
\begin{CJK*}{UTF8}{gbsn}``walking for cash (走路赚钱)''\end{CJK*}---a reference to a famous Chinese ``step for cash'' mobile fitness application \begin{CJK*}{UTF8}{gbsn}QuBu (趣步)\end{CJK*}, which allowed users to earn cash by achieving a certain amount of steps and inviting friends. 
English participants also described similar experiences, with one user reporting: ``I felt like I had to check the app to keep up.''




\section{RQ\#2: Emotions Related to Manipulative Experiences}
\label{sec:emotion}
\begin{figure}[h]
\subfloat[\textit{Emotional Response}]{{\includegraphics[width=0.45\textwidth]{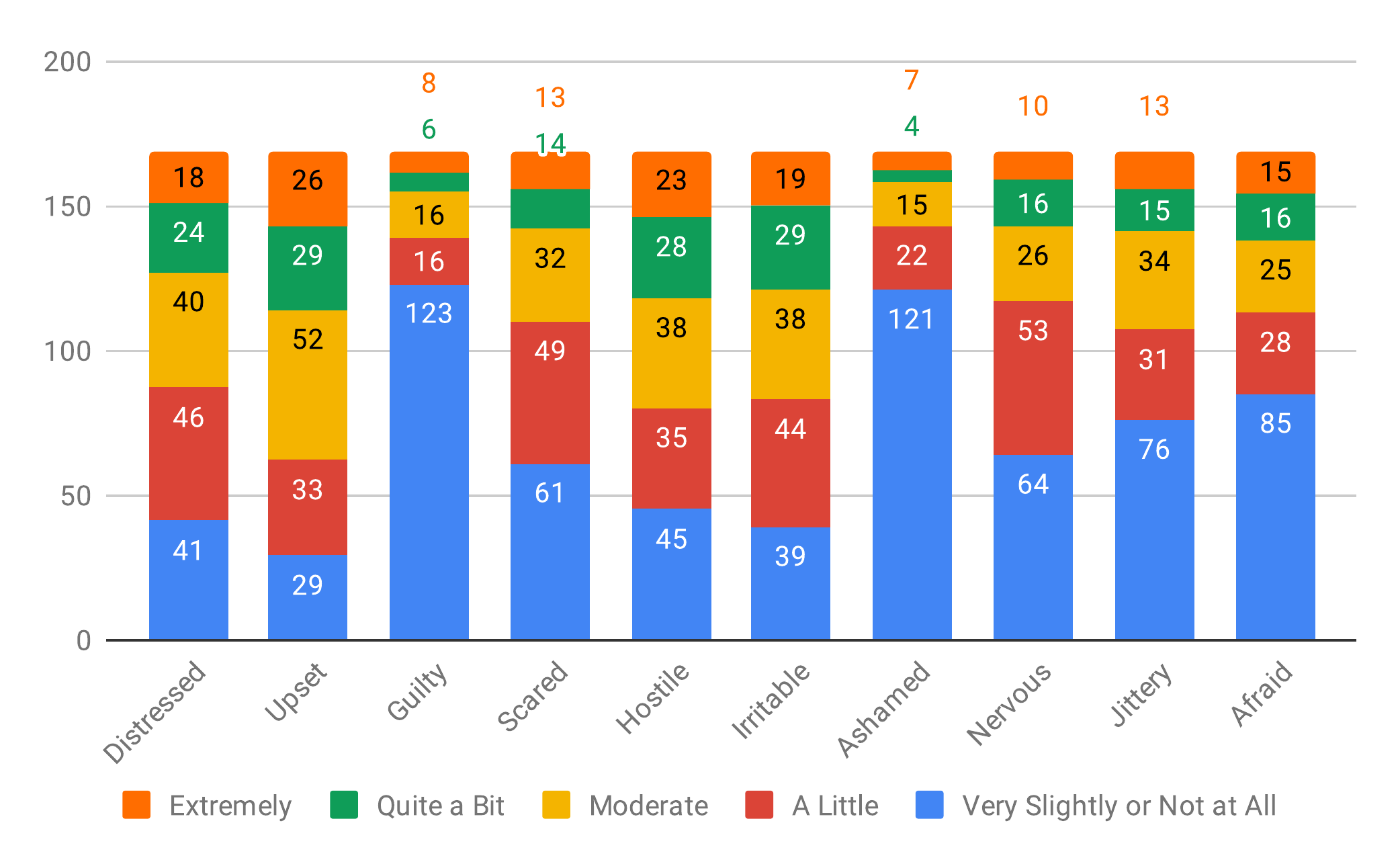} }}%
    \qquad
    \subfloat[\textit{Mostly felt emotions vs. Age}]{{\includegraphics[width=0.45\textwidth]{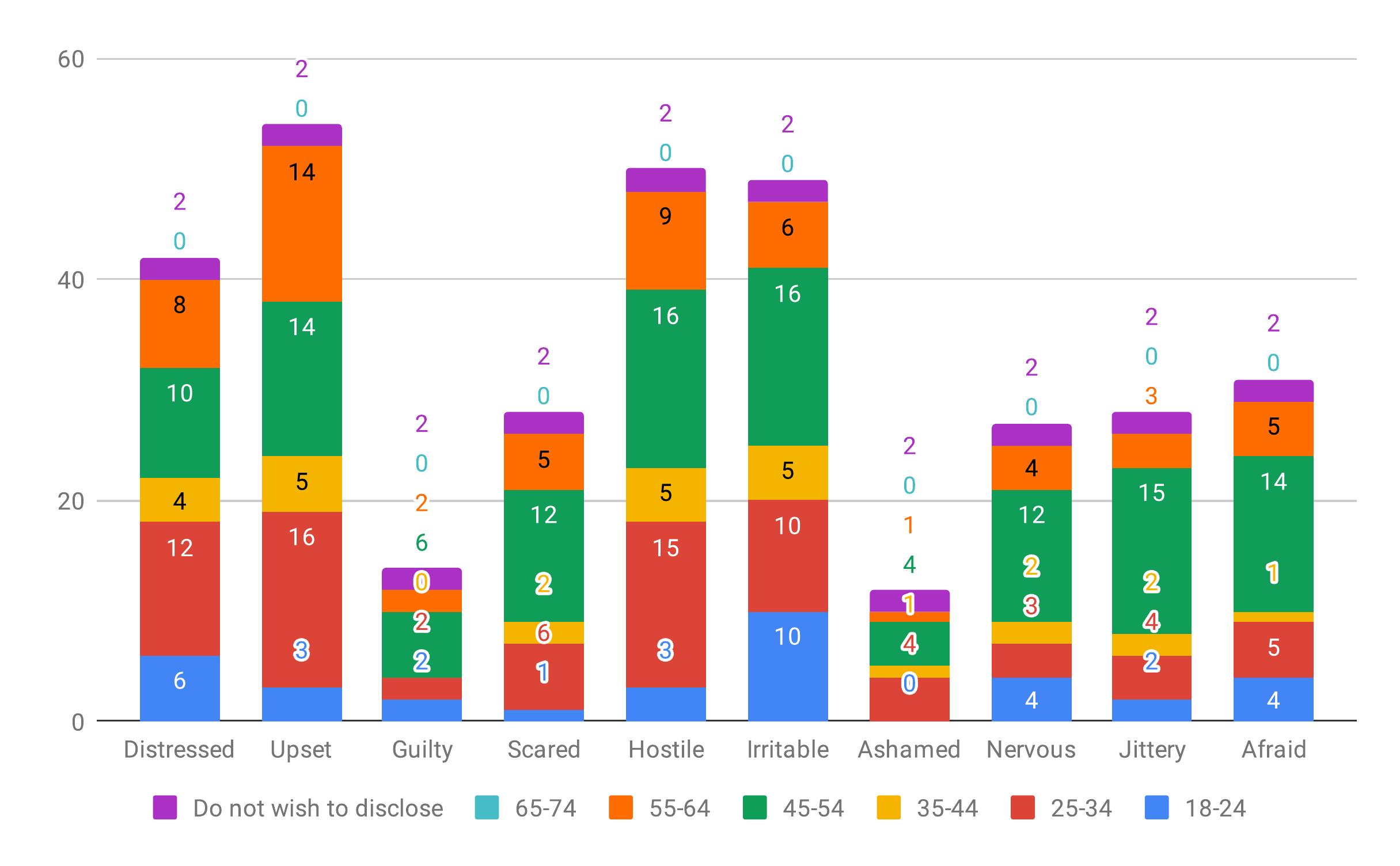} }}%
    \caption{Users' Emotional Response to Manipulation.}%
    \label{emotion}%
    \end{figure}
We asked users to rank their felt emotions in relation to past manipulative digital product experiences on a scale from ``Very Slightly or Not at all'' to ``Extremely'' using a list of the following negative emotions: Distressed, Upset, Guilty, Scared, Hostile, Irritable, Ashamed, Nervous, Jittery, and Afraid; illustrated in Figure~\ref{emotion}(a). Users frequently expressed that they felt strong emotions such as being distressed (n=82), upset(n=107), hostile (n=89), and irritable (n=86) when they experienced manipulative products. Users also responded in a negative but more neutral emotions of being scared (n=59), nervous (n=52), jittery (n=62), and afraid (n=56) through the manipulative experience.
Users responded least strongly (``Very slightly or Not at all'') to feeling guilty (n=123) or ashamed (n=121) through their manipulative experience, which resonated with a smaller number of users (n=34) blaming themselves for the manipulation (results detailed in Section~\ref{sec:awareness}). In Figure~\ref{emotion}(b), we have illustrated the frequency of commonly felt emotions (adding responses for ``Quite a bit'' and ``Extremely'') as represented across different age groups. The bar heights represent that the number of respondents under each age group were uneven, but these initial non-normalized descriptive statistics show how users emotionally responded to felt manipulation, which may point to interesting areas for future work. Within each age group, we observed a somewhat different proportion of users feeling irritable, hostile, or upset; for example, the proportion of users expressing irritability in the age group 18-24 yrs was 55.6\% (n=10) while for the age group 45-54 yrs, 27.8\% (n=16) of users expressed this emotion. 

\section{RQ\#3: Users' Awareness about the Creation and Creators of Manipulation Products}
\begin{figure}[H]
    \centering
    \includegraphics[width=0.5\textwidth]{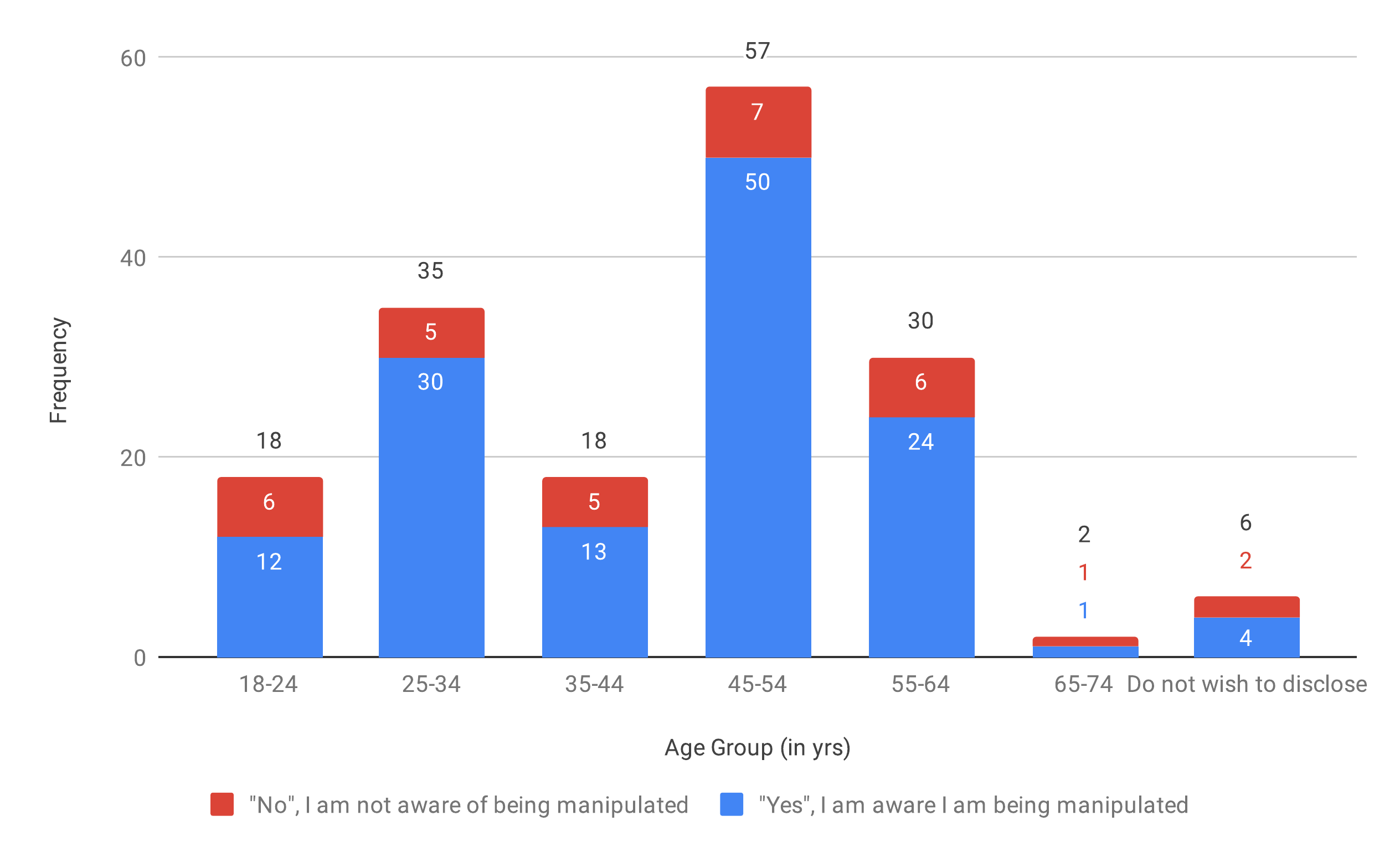}
    \caption{User's Awareness of Being Manipulated by Age Group.}
    \label{awarenessage}
\end{figure}

\subsection{Level of Awareness}
\label{sec:awareness}
In this section, we describe the users' awareness of being manipulated, including who they blame or feel is responsible for creating manipulative products, and how often they mistrust the smartphone applications or website they regularly interacting with. When asked if users felt that smartphone applications or websites were designed to manipulate them, 79.3\% of the users were aware of being manipulated (`Yes': n= 134; `No': n=32; n=3 did not respond). The overall proportion of users responding to this question was relatively stable by age, with more than two-thirds of respondents in each age bracket responding in the affirmative~\ref{awarenessage}. 

\begin{figure}[h]
\subfloat[\textit{"Who do you think is to blame?"}]{{\includegraphics[width=0.55\textwidth]{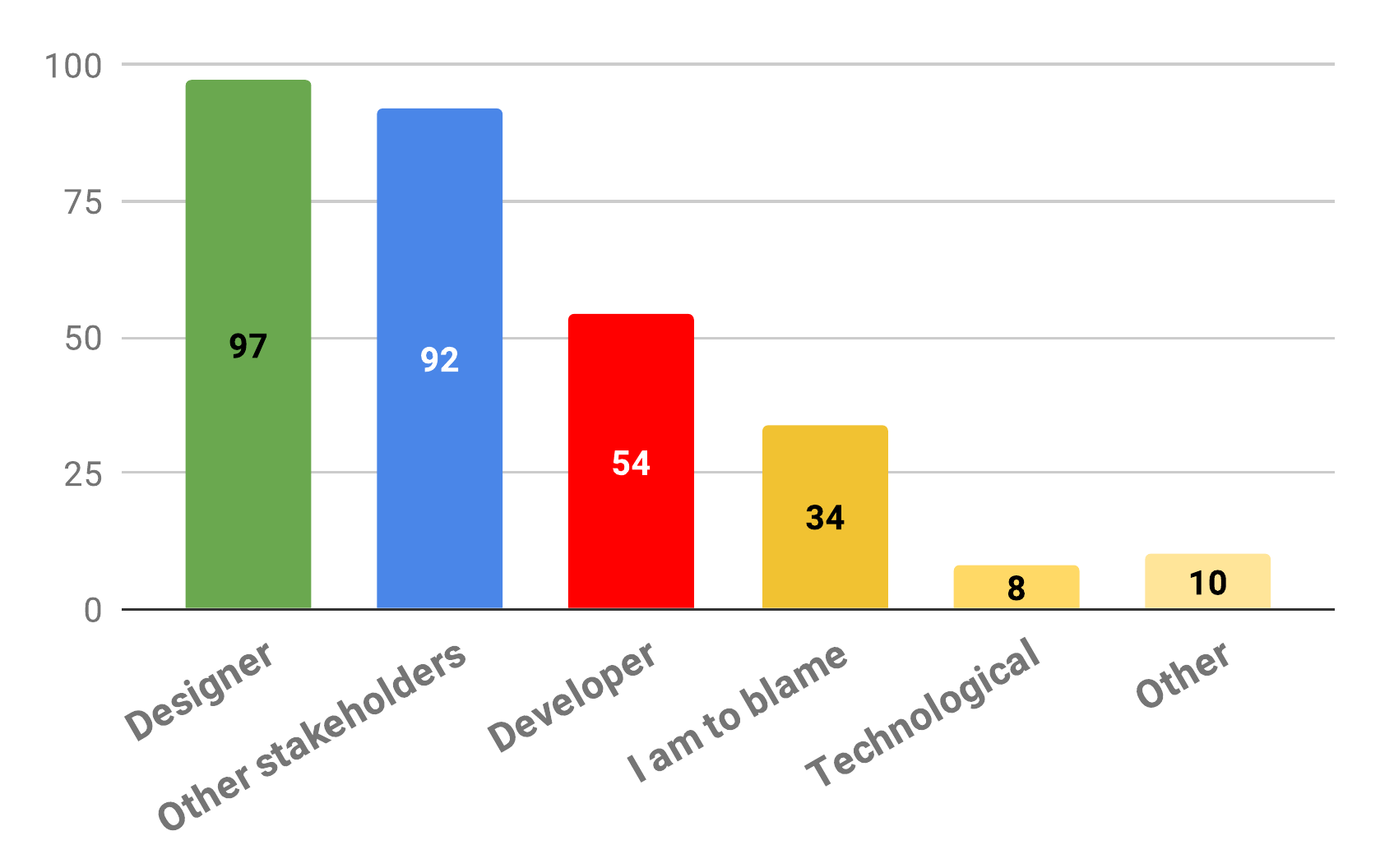} }}%
    \qquad
    \subfloat[\textit{Who among the creators is to blame?}]{{\includegraphics[width=0.35\textwidth]{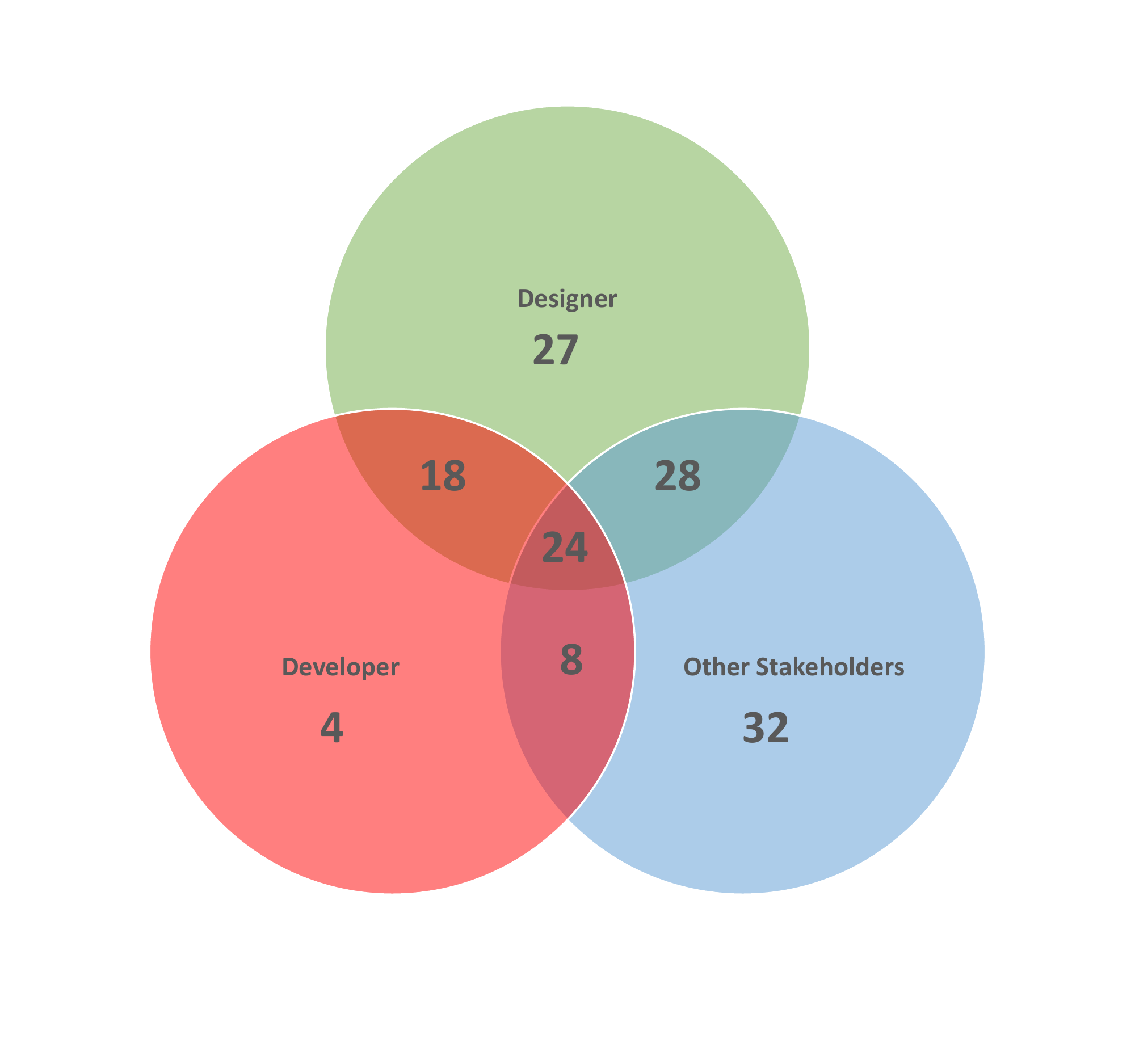} }}%
    \caption{Users' Awareness of the Creators of Manipulative Interfaces.}%
    \label{blamegame}%
\end{figure}
    
We also asked the users in a multi-option question who they felt was to blame for the manipulation they experienced in digital products; they could select one or more items from a list including 1)``I am to blame, as the user''; 2) The designer of the site or application (e.g., the designer created the application or website to trick me into doing something.); 3) The developer of the site or application (e.g., the developer created the application or website to trick me into doing something.); 4) Other stakeholders (e.g., a company, the company wants to trick me into doing something); 5) Technological Issues (e.g., screen or device quality); and 6) Other, with optional textbox to add to the list. As shown in Figure~\ref{blamegame}(a), most of the users blamed the designer (n=97), other stakeholders (n=92) and developer (n=54) for creating manipulative applications or websites intended to trick them. A minority blamed themselves as the user (n=34), and users rarely blamed feelings of manipulation on technological issues (n=8). Open-ended responses (n=10) included statements such as: ``\textit{The owner of the application and/or advertiser}'' and ``\textit{\ldots it’s not intended to manipulate and has a justifiable reason, but it isn’t communicated well}.'' To demonstrate the overlaps among these targets of manipulation, we present a Venn diagram in Figure~\ref{blamegame}(b) to describe how users blamed the varying creators (designers, developers and other stakeholders) of smartphone applications and websites. In these overlapping responses, 14.2\% (n=24) of the users blamed all of the creators of these digital products---designers, developers and other stakeholders; 71.6\% of the users blamed only designers and company stakeholders (n=121); and the least number of users (n=28) blamed only the developer of the application and website. 


\begin{figure}[h]
    \includegraphics[width=0.45\textwidth]{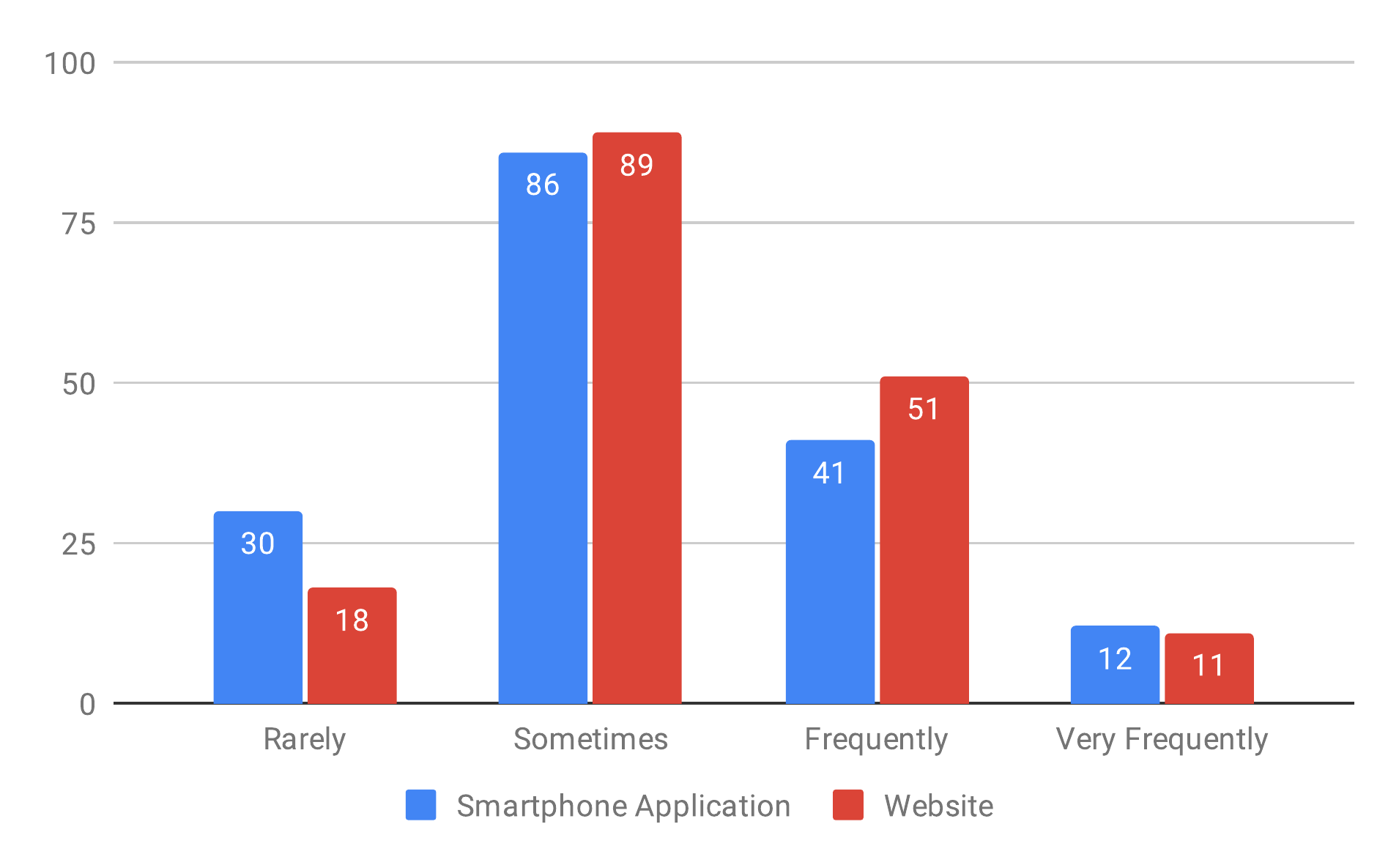}
    \caption{Users' Response to Mistrust of Smartphone Applications or Websites.}%
    \label{mistrust}%
\end{figure}
    
Finally, we asked the users how often they mistrust the smartphone applications and websites they regularly interact with. In Figure~\ref{mistrust}, we present the frequencies of these responses on a scale of ``rarely,'' ``sometimes,'' ``frequently,'' and ``very frequently.'' For smartphone applications, 17.8\% (n=30) of the users rarely felt mistrust while 31.4\% (n=53) of the users frequently mistrusted these applications. Similarly for websites, 10.7\% (n=18) of the users rarely felt mistrust and 36.7\% (n=62) users frequently mistrusted the websites. Although the difference is minimal, users appeared to mistrust websites more commonly than smartphone applications. A majority of users (n=86 for smartphone application; n=89 for website) identified that they mistrusted their interactions with these digital products ``sometimes,'' ad combining the results of ``sometimes,'' ``frequently,'' and ``very frequently'' yielded a majority of responses for both smartphone applications (82.24\%; n=139) and websites (89.3\%; n=151). 

\subsection{Creators of Manipulative Digital Products}
\label{Sec:creators}
\begin{figure}[h]
\subfloat[\textit{How all users they were felt they were valued?}]{{\includegraphics[width=0.45\textwidth]{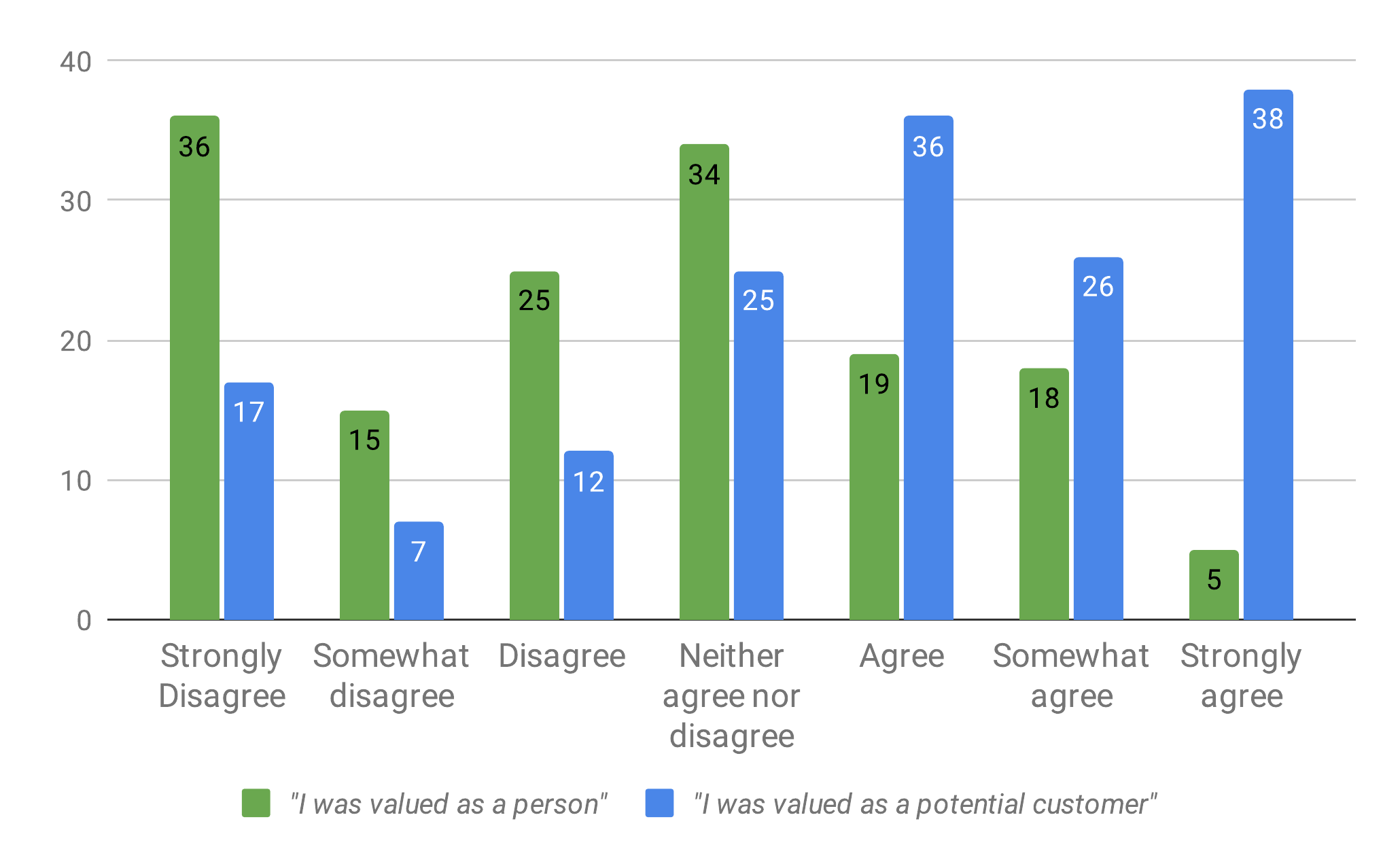} }}%
    \qquad
    \subfloat[\textit{How users not aware of manipulation felt they were valued?}]{{\includegraphics[width=0.45\textwidth]{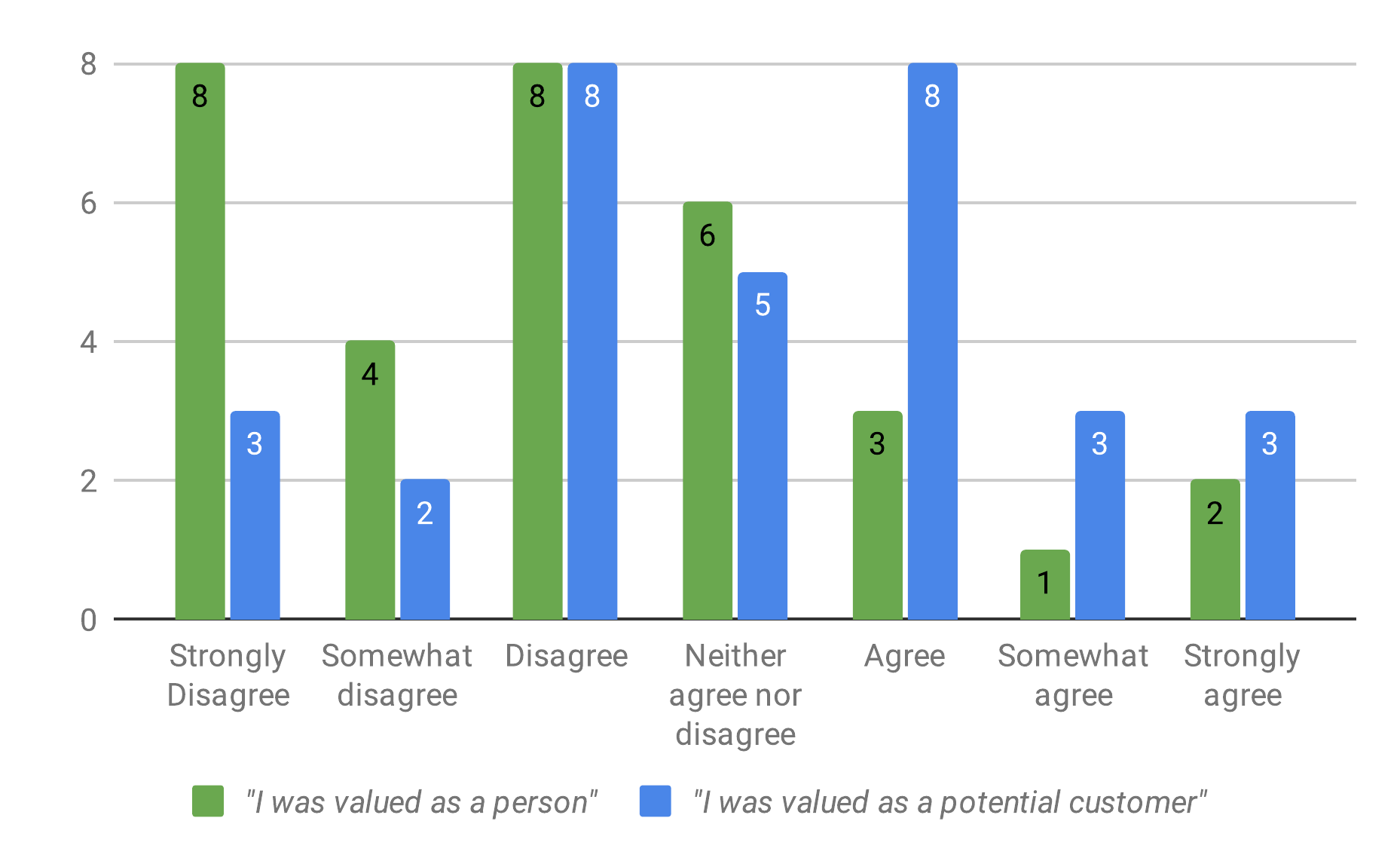} }}%
    \caption{Users' Sense of How They Were Valued.}%
    \label{value}%
    \end{figure}
    
We asked the users how they felt they were valued by the creators of smartphone applications and websites, using a scale from ``strongly disagree'' to ``strongly agree'' to two respond to two statements: 1) ``I was valued as a person'' and 2) ``I was valued as a customer.'' As shown in Figure~\ref{value}(a), 62.11\% of the users (n=100) felt they were valued as a customer with responses of ``agree'' or higher, with only 27.6\% (n=42) feeling they were valued as a person with responses of ``agree'' or  higher. 50\% of the respondents (n=76) responded with disagree or lower to ``I was valued as a person.'' 
To further explore this positioning as a person or as a customer, we also identified how users responded to these two questions if they had previously identified that they were not aware of being manipulated (n=32) in Figure~\ref{value}(b). Even if participants were not aware of the manipulation, they still felt that they were valued more as a customer than a person, representing what appears to be a broad consensus that the creators of digital products treat the users as consumers despite the presence or awareness of explicit manipulative techniques.

\subsection{User Perceptions of Digital Product Creators}
In this final section, we present vignettes from our follow-up  interviews with users in order to elaborate on users' awareness about the creators of manipulative digital products and relate these experiences to perceptions of manipulation and the progression of these feelings over time (Section~\ref{sec:continuum}). We present three pseudonymous vignettes below, including Amy (English), Liu (Chinese), and Wang (Chinese).


\subsubsection{Vignette 1: Trapped in Auto-Subscription}

Amy began her interaction with a digital product 
by clicking on a Facebook post about a face cream that she later realized was ``too good to be true.'' At the time, this interaction did not cause her to be aware of any explicit manipulation, but after realizing that she had been auto-subscribed, Amy felt upset because she didn't ``know everything in the beginning.'' Upon later reflection, she felt that the creators of this interaction set a trap by hiding necessary information, not allowing her to make the smartest decision. This auto-subscription ended up costing Amy both money and time---identifying that the face cream auto-renewed, and then working to cancel the subscription. Because she did not sense the manipulative intent in her initial ``inspection of details,'' the consequences of being manipulated were only experienced after felt ``negative results from interaction.'' 

After this negative experience, she blamed herself as a user, but also cast blame on the designers of the interaction and other stakeholders responsible for the negative results. She blamed herself for not being smart enough to identify the ``trap'' and in the future, told us that she would ignore similar social media posts or advertisements that felt ``too good to be true'' to avoid being ``in the same situation.'' Relating her experience to the perceived creators of this interaction, she believed that ``designers have total control over what [users] see'' and as creators, ``they have got pressure on them to produce customers, so they have to do whatever they can to make it work.'' With these business goals in mind, ``\textit{[the designers] are not thinking about others as much as they are thinking of themselves}.'' This manipulation intent---and the justification she felt from the designers' perspective---made her feel that the creators were ``valuing [her] for her money but not for being a person.'' She concluded that these creators don't have ``strong values about right and wrong'' and if they had a strong will to serve their customers, ``(these creators) couldn't stay in [their] job if it was always going to be about money and survival.'' Amy clearly expressed that the creators had to trade off user and social values for business and monetary benefit. 



 \subsubsection{Vignette 2: Gamified ``Stepping for Cash''}
Liu has long recognized the gamification of various components in the QuBu fitness application that she uses to ``step for cash.'' Initially, she thought Qubu was trustworthy because it was recommended by her friends, but after using the app, Liu discovered she had to \begin{CJK*}{UTF8}{gbsn}``发展下线 (market the app to more people)''\end{CJK*} to keep up with her progress and earn more \begin{CJK*}{UTF8}{gbsn} ``糖果 (candies---a digital currency within QuBu, which could be used to exchange cash or products).''\end{CJK*} Later, Liu stopped using QuBu since she was \begin{CJK*}{UTF8}{gbsn}``恐惧 (fearful)''\end{CJK*} that being \begin{CJK*}{UTF8}{gbsn}``过多的参与进去 (too involved into [QuBu])'' might get her into trouble\end{CJK*}, and she concluded that Qubu wasn't simply a fitness app as they advertised: it was \begin{CJK*}{UTF8}{gbsn}``变相的传销 (a variation of multi-level-marketing, which is explicitly prohibited by the Chinese government.)''\end{CJK*} Similar to Amy, Liu did not sense the manipulative intent in her ``general conclusion'' early on, and the consequences of being manipulated were only experienced after extended interaction, including ``undesired interactions'' and ``negative results from interaction.'' 

She expressed strong feelings toward Qubu's creators, saying that its creators are \begin{CJK*}{UTF8}{gbsn} ``令人深恶痛绝的 (abhorrent)'' and she mentioned it was ``(可恶) hateful''\end{CJK*} when designers and stakeholders make business decisions based primarily on \begin{CJK*}{UTF8}{gbsn}``经济利益 (financial incentives)''. She stated that manipulative technology creators are ``[道德]概念很模糊 (ambiguous about ethics)''，and ``急功近利的追求财富 ([these creators] would like to do whatever they can to pursue financial success in a short time frame.)\end{CJK*} '' Therefore, they were willing to enter ``gray areas'' of the law and take advantages of spaces where  \begin{CJK*}{UTF8}{gbsn}``中国的很多制度还不健全 (there isn't a lot of regulation in China [for technology manipulation.])''\end{CJK*} On the other hand,  Liu blamed herself for not being careful---because it was she who wanted to use the application and \begin{CJK*}{UTF8}{gbsn}``想赚钱 (make money [through Qubu])''\end{CJK*} in the first place. She elaborates this assessment, noting that the creators of QuBu \begin{CJK*}{UTF8}{gbsn}``没偷没抢 (did not break any laws)''\end{CJK*}; therefore, Liu had trouble in assigning full blame to Qubu's creators for manipulating her because \begin{CJK*}{UTF8}{gbsn}``
很难界定(it was hard to decide from what perspective to evaluate [them.])''\end{CJK*} 

 \subsubsection{Vignette 3: Forced Collection of Personal Information}
Wang has been concerned about enabling phone permissions for years due to his fears of being tracked and potentially manipulated by others. He shared an experience about his use of a vehicle rental application, where he had to allow several smartphone permissions that he felt were unnecessary. Wang thought about using a similar application as an alternative; however, in that situation, the vehicle rental app that required these permissions was his only option. Therefore, Wang felt he was \begin{CJK*}{UTF8}{gbsn}``被逼无奈 (helpless and forced to [enable permissions without a choice.])''\end{CJK*} Wang further expressed that he was frequently concerned about apps having access to his personal data, because he worried about \begin{CJK*}{UTF8}{gbsn}``信息泄露 ([apps] leaking his information)''\end{CJK*} and \begin{CJK*}{UTF8}{gbsn}``有人把它收集起来对我做个画像...那么可能就会有损我的个人信息安全 (intentional data collection, which can be used to create a behavior model and harm my personal information security.)''\end{CJK*} Because Wang sensed the manipulative intent through an ``undesired interaction,'' his feelings of being manipulated were heightened, even though he had no option to use a competing app or service. 

Wang blamed technology designers and stakeholders for this experience that he felt was manipulative. He pointed toward the issue of the creator's ethical awareness, further illustrating his perception by citing a Chinese proverb: \begin{CJK*}{UTF8}{gbsn} ``不为五斗米折腰 (don't curtsy for the salary of five bushels of rice; meaning that people should not lose their dignity because of money and do something that's ethically wrong.)''\end{CJK*} He felt that these creators were \begin{CJK*}{UTF8}{gbsn}``短视的 (short sighted)''\end{CJK*} because they focused on \begin{CJK*}{UTF8}{gbsn}``蒙一下是一下 (tricking as many people as they can)''\end{CJK*} as an exchange of monetary incentives. If these creators wanted to build an \begin{CJK*}{UTF8}{gbsn} ``百年老店 (evergreen brand)''\end{CJK*}, Wang claimed that the company would be able to \begin{CJK*}{UTF8}{gbsn}``把服务持续的提供给大家 (provide [high-quality] services to customers consistently.)''\end{CJK*} Additionally, compared with the ethical responsibility of designers, Wang thought that stakeholders should be more responsible to disincentive technology manipulation. He believed that \begin{CJK*}{UTF8}{gbsn}``设计师想把产品做的更好 (designers wanted to create better products)''\end{CJK*}, but they could \begin{CJK*}{UTF8}{gbsn}``受资本的驱动 (be under the stress of capitalists.)''\end{CJK*} In conclusion, he felt that designers and developers had the responsibility to \begin{CJK*}{UTF8}{gbsn}``平衡 (balance)''\end{CJK*} stakeholders' desires and users' rights to reach \begin{CJK*}{UTF8}{gbsn}``互相说服 (mutual agreement.)''\end{CJK*}


\section{Discussion}

\subsection{Linking Experiences of Manipulation to Dark Patterns Strategies}
While we did not directly ask participants about their experiences with ``dark patterns''---we did use ``felt manipulation'' as a proxy to investigate how dark-patterns-informed digital products might be experienced by end users. In this subsection, we seek to link our findings to broader notions of dark patterns in the literature, noting where dark patterns strategies align to users' perceptions of felt manipulation and experiences of manipulation over time.

First, we draw together the perceptions of manipulation we shared in Section~\ref{sec:themes} with dark patterns literature and the previous end user findings of Maier and Harr~\cite{Maier2020-yk} (e.g., notions of perception that include impressions, assessment, balance, and acceptability). The majority of the issues that drove users' perception of felt manipulation included spaces where users had to engage in an \textit{assessment} of a context or experience, including the connection of broader themes of past personal and societal experiences and the desire for personal control. The examples we have collected from participants demonstrates this ongoing \textit{assessment} of digital experiences and \textit{balancing} of value versus the perceived threat to decide whether to more fully engage and use the digital product. Threats that emerged from the users' assessment included data collection and use at personal and ``big data'' scales (collection of personal information, threats of big data); an overall perception that resulted from a contextually aware judgment that included context, user experiences or reviews, and degree of familiarity (overall perception of trust and distrust); and the use of products whose functionality and risk assessment may change over time due to monetization strategies (use of freemium products). The users' engagement with different forms of \textit{assessment} and \textit{balancing} of user needs and digital product capabilities and risks builds upon these end user perceptions as defined by Maier and Harr~\cite{Maier2020-yk}, while also providing additional points of departure by which these assessments and decisions might be made.

Second, we compare the emergence of felt manipulation we shared in Section~\ref{sec:continuum} with known dark patterns strategies from Gray et al.~\cite{Gray2018-or} to better identify where awareness of different strategies might emerge along this temporal progression. Based on our card sort, we identified 24 instances of felt manipulation as an initial judgment, while 130 instances of felt manipulation occurred after initial judgment but prior to any user interaction (including inspection of details, felt persuasion, and general conclusion). This finding points towards broad awareness that something is ``off'' or ``not correct,'' but often lacking more precise language to describe what is driving the feeling that the user is being manipulated. This aligns well with dark patterns strategies such as \textit{sneaking} and \textit{interface interference}, where user behaviors are being nudged, but in ways that are intended to be largely undetectable without further investigation. Other dark patterns strategies such as \textit{obstruction} or \textit{forced action} might only be detected as potentially manipulative once a user engages in an interaction and receives an negative result (n=24) or when a user experiences an undesired or unnecessary interaction (n=30). Similarly, the dark pattern strategy of \textit{nagging} could only be experienced over time, resulting in impacts on only the latter half of the manipulative continuum~(Table~\ref{continuum}). What this analysis and comparison with dark patterns strategies begins to reveal in greater detail is the role of temporality, awareness, and translation of emotions to designer intent that occurs rapidly in an end user's engagement with technology. While end users may not have sufficient language to describe manipulation in an overarching and conceptual sense---such as the driving language of ``dark patterns''---they nevertheless are able to identify a range of issues that cause them to suspect manipulative intent, and can share examples of this manipulation that extend across the entire temporal user journey. This finding aligns with Maier and Harr~\cite{Maier2020-yk}, positioning the user impressions, assessment, and identification of acceptability as key factors in deciding whether to use a system (if such a choice is possible), or whether detected manipulative intent warrants further scrutiny, care, or protection.

\subsection{Felt Manipulation as a Potential Platform for User Advocacy}
So what should we do about the finding that 82.24\% of users mistrusted smartphone applications and 89.3\% of users mistrusted websites at least ``sometimes''? This finding aligns with Di Geronimo et al. \cite{Di_Geronimo2020-mh} in the context of smartphone applications, where they found that 95\% of the apps they evaluated exhibited the presence of one or more dark patterns---but what recourse do users have, even if they are able to monitor their experiences and identify sources and points of manipulation? We propose the use of our initial frameworks of manipulative perception and temporal progression of felt manipulation to be an important first step in empowering users to take control of their digital experiences. To name something---assigning it the label of ``manipulative,'' ``dark,'' ``sneaky,'' or downright evil---is an important first step of awareness that is necessary to lead to future advocacy efforts. While many of the techniques that users identified are not strictly illegal in most regions of the world, these problematic practices can be made illegal with the efforts of a wide range of individuals, including users themselves. As has become clear in the wake of GDPR, end users can be given rights that allow them to exercise control over parts of their digital experience (cf., Recital 59~\cite{recital59}). While many of the processes by which users can exercise these rights are still under active construction or negotiation, this might provide one way to envision the ability of users to collectively identify and characterize certain digital products---or strategies contained within digital products---as suspect, potentially manipulative, or illegal, knowing that many companies might not have interest in changing their practices on their own. 
While we have identified multiple sources of ``felt manipulation,'' advocacy might be taken up along multiple different lines. One possible direction is simply the ``mainstreamification'' of the term ``dark patterns,'' allowing users to co-opt the evolving language proposed by a range of HCI and design scholars~\cite{Narayanan2020-va,Gray2018-or,Chromik2019-rm,Greenberg2014-dg,Mathur2019-ea}. This strategy may succeed along some dimensions, raising awareness of manipulative strategies built into technologies, while also perhaps leading to a watering down of what is truly a \textit{psychological trick} \cite{Brignull2013-ap} or an \textit{imbalance of shareholder and user value} \cite{Gray2018-or}. Even with crowdsourced additions to sites that collect dark patterns exemplars (@darkpatterns on Twitter\footnote{https://twitter.com/darkpatterns}; a searchable set of examples from Gray and colleagues\footnote{https://darkpatterns.uxp2.com/patterns/}; discussions the ``r/assholedesign'' subreddit\footnote{https://www.reddit.com/r/assholedesign/} \cite{Gray2020-zq}; a community to find ``healthy mobile games''\footnote{https://www.darkpattern.games}), it is yet unclear whether merely collecting and voting on manipulative exemplars necessarily leads to feelings of empowerment and advocacy. Another possible direction may be a more lightweight means of awareness, perhaps focusing on different set of language that may be taken up more colloquially---perhaps as an instantiation of a mental model of persuasion, addiction, or manipulation that is increasingly part of the popular discourse around technology ethics (cf., the problematic yet popular documentary \textit{The Social Dilemma}\footnote{https://www.thesocialdilemma.com}). A shift around language might foreground capitalist goals as being focused on persuasion (in the best of cases), manipulation (in typical cases), or coercion (in cases where there are no alternative options), allowing users a more expanded vocabulary to link their feelings of unease with possible design intentions. 

\section{Future Work}
Our findings and discussion point towards numerous areas for future work, which we can only outline in the broadest possible way here. Because the intersection of dark patterns, design intent, and end user perceptions has been historically understudied, there are numerous research directions that could focus on: 1) characterizing end user emotions and sensitivity when perceiving ``dark'' UX experiences; 2) identifying factors that contribute to less or more sensitivity to manipulative strategies through broad-scale experimental and survey research, including factors such as age, profession, technology experience, and ethical training; 3) means of qualitatively assessing and identifying through user testing or evaluation as part of a product lifecycle; 4) describing specific aspects of felt manipulation both theoretically and conceptually; 5) identifying opportunities for advocacy in identifying problematic strategies and contributing to policy development, potentially through participatory or crowdsourced means; 6) assessing and creating new supports for design students and practitioners to better understand how and where felt manipulative intent might emerge; and 7) a convergent and transdisciplinary description of what factors are likely to contribute to a holistic assessment of manipulation from multiple disciplinary and stakeholder perspectives.

To demonstrate these opportunities for future work further, we will elaborate a couple of these directions further. For example, more research on describing specific facets of felt manipulation could build on our findings and the work of Di Geronimo et al. \cite{Di_Geronimo2020-mh}, based on the emergent insight that greater user sensitivity to the \textit{presence} of dark patterns may lead to greater success in users detecting patterns in future digital product experiences. Similar and more extensive experimental assessments of specific dark patterns across specific moments in the perception continuum ---from initial judgment to extended interaction---might reveal spaces where users are already well primed to identify manipulative practices, perhaps leveraging the early experimental work of Luguri and Strahilevitz \cite{Luguri2019-bg} that focuses on sensitivity and potential harm of individual dark patterns strategies. As an alternate example, future work on connecting user awareness with public policy and governance may facilitate the translation of user perception to actionable legal recourse, similar to the data processing governance structure currently in place to support GDPR. Work in this area might combine sensitization towards manipulative experiences, documented above, with platforms to organize as a collective action campaign, potentially using insights from digital civics, participatory design, and crowdwork to identify problematic strategies or products and argue for their change or discontinuance. 

\section{Conclusion}
In this paper, we have reported on the results of a survey and followup interviews conducted with English and Mandarin Chinese-speaking participants, further documenting the felt characteristics of digital product experiences that relate to manipulation and the presence of dark patterns. We described a range of perception characteristics that lead to an assessment of manipulation, including a temporal progression of these experiential characteristics that reinforce a judgment of manipulation across initial perception through extended interaction. We built upon these perceptions of manipulation to reveal common emotions experienced by end users, and the ways in which these users described the designer intent of these systems and the individuals they hold responsible for manipulative experiences. Based on these findings, we offer additional opportunities to intersect research on dark patterns and policy with these assessments of manipulation, providing new avenues for supporting user advocacy and agency.

\begin{acks}
We gratefully acknowledge the efforts of Madison Fansher in developing a pilot version of the survey, later expanded with the help of undergraduate researcher Lucca McKay. We also thank Zexi (Jessie) Zhou for her help in translating the survey and interviews. This work is funded in part by the National Science Foundation under Grant Nos. 1657310 and 1909714. 
\end{acks}

\bibliographystyle{ACM-Reference-Format}
\bibliography{sample-base}


\begin{thebibliography}{57}


\ifx \showCODEN    \undefined \def \showCODEN     #1{\unskip}     \fi
\ifx \showDOI      \undefined \def \showDOI       #1{#1}\fi
\ifx \showISBNx    \undefined \def \showISBNx     #1{\unskip}     \fi
\ifx \showISBNxiii \undefined \def \showISBNxiii  #1{\unskip}     \fi
\ifx \showISSN     \undefined \def \showISSN      #1{\unskip}     \fi
\ifx \showLCCN     \undefined \def \showLCCN      #1{\unskip}     \fi
\ifx \shownote     \undefined \def \shownote      #1{#1}          \fi
\ifx \showarticletitle \undefined \def \showarticletitle #1{#1}   \fi
\ifx \showURL      \undefined \def \showURL       {\relax}        \fi
\providecommand\bibfield[2]{#2}
\providecommand\bibinfo[2]{#2}
\providecommand\natexlab[1]{#1}
\providecommand\showeprint[2][]{arXiv:#2}

\bibitem[\protect\citeauthoryear{??}{fee}{[n.d.]}]%
        {feelingwheel}
 \bibinfo{year}{[n.d.]}\natexlab{}.
\newblock \bibinfo{title}{Feelings Wheel}.
\newblock \bibinfo{howpublished}{\url{http://feelingswheel.com/}}.
\newblock
\newblock
\shownote{Accessed: 2020-10-13.}


\bibitem[\protect\citeauthoryear{??}{fir}{[n.d.]}]%
        {firstthingsfirst2000}
 \bibinfo{year}{[n.d.]}\natexlab{}.
\newblock \bibinfo{title}{First Things First Manifesto 2000}.
\newblock
  \bibinfo{howpublished}{\url{http://www.eyemagazine.com/feature/article/first-things-first-manifesto-2000}}.
\newblock
\urldef\tempurl%
\url{http://www.eyemagazine.com/feature/article/first-things-first-manifesto-2000}
\showURL{%
\tempurl}
\newblock
\shownote{Accessed: 2020-10-10.}


\bibitem[\protect\citeauthoryear{??}{noa}{[n.d.]}]%
        {noauthor_undated-gy}
 \bibinfo{year}{[n.d.]}\natexlab{}.
\newblock \bibinfo{title}{General Data Protection Regulation ({GDPR}) --
  Official Legal Text}.
\newblock \bibinfo{howpublished}{\url{https://gdpr-info.eu/}}.
\newblock
\urldef\tempurl%
\url{https://gdpr-info.eu/}
\showURL{%
\tempurl}
\newblock
\shownote{Accessed: 2020-9-19.}


\bibitem[\protect\citeauthoryear{??}{rec}{[n.d.]}]%
        {recital59}
 \bibinfo{year}{[n.d.]}\natexlab{}.
\newblock \bibinfo{title}{Recital 59 - Procedures for the Exercise of the
  Rights of the Data Subjects}.
\newblock \bibinfo{howpublished}{\url{https://gdpr-info.eu/recitals/no-59/}}.
\newblock
\urldef\tempurl%
\url{https://gdpr-info.eu/recitals/no-59/}
\showURL{%
\tempurl}
\newblock
\shownote{Accessed: 2020-10-15.}


\bibitem[\protect\citeauthoryear{??}{noa}{2018}]%
        {noauthor_2018-ww}
 \bibinfo{year}{2018}\natexlab{}.
\newblock \bibinfo{title}{California Consumer Privacy Act ({CCPA})}.
\newblock \bibinfo{howpublished}{\url{https://oag.ca.gov/privacy/ccpa}}.
\newblock
\urldef\tempurl%
\url{https://oag.ca.gov/privacy/ccpa}
\showURL{%
\tempurl}
\newblock
\shownote{Accessed: 2020-9-19.}


\bibitem[\protect\citeauthoryear{Ackerman}{Ackerman}{1995}]%
        {ackerman1995concept}
\bibfield{author}{\bibinfo{person}{Felicia Ackerman}.}
  \bibinfo{year}{1995}\natexlab{}.
\newblock \showarticletitle{The concept of manipulativeness}.
\newblock \bibinfo{journal}{\emph{Philosophical Perspectives}}
  \bibinfo{volume}{9} (\bibinfo{year}{1995}), \bibinfo{pages}{335--340}.
\newblock


\bibitem[\protect\citeauthoryear{Bohnsack and Liesner}{Bohnsack and
  Liesner}{2019}]%
        {Bohnsack2019-py}
\bibfield{author}{\bibinfo{person}{Ren{\'e} Bohnsack} {and}
  \bibinfo{person}{Meike~Malena Liesner}.} \bibinfo{year}{2019}\natexlab{}.
\newblock \showarticletitle{What the hack? A growth hacking taxonomy and
  practical applications for firms}.
\newblock \bibinfo{journal}{\emph{Business horizons}} \bibinfo{volume}{62},
  \bibinfo{number}{6} (\bibinfo{date}{Nov.} \bibinfo{year}{2019}),
  \bibinfo{pages}{799--818}.
\newblock
\showISSN{0007-6813}
\urldef\tempurl%
\url{https://doi.org/10.1016/j.bushor.2019.09.001}
\showDOI{\tempurl}


\bibitem[\protect\citeauthoryear{Bowles}{Bowles}{2018}]%
        {Bowles2018-gx}
\bibfield{author}{\bibinfo{person}{Cennydd Bowles}.}
  \bibinfo{year}{2018}\natexlab{}.
\newblock \bibinfo{booktitle}{\emph{Future Ethics}}.
\newblock \bibinfo{publisher}{Nownext Press}.
\newblock
\showISBNx{9781999601911}
\urldef\tempurl%
\url{https://market.android.com/details?id=book-fAy_vAEACAAJ}
\showURL{%
\tempurl}


\bibitem[\protect\citeauthoryear{Brignull}{Brignull}{2011}]%
        {Brignull2011-id}
\bibfield{author}{\bibinfo{person}{Harry Brignull}.}
  \bibinfo{year}{2011}\natexlab{}.
\newblock \showarticletitle{Dark Patterns: Deception vs. Honesty in {UI}
  Design}.
\newblock \bibinfo{journal}{\emph{Interaction Design, Usability}}
  \bibinfo{volume}{338} (\bibinfo{year}{2011}).
\newblock


\bibitem[\protect\citeauthoryear{Brignull}{Brignull}{2013}]%
        {Brignull2013-ap}
\bibfield{author}{\bibinfo{person}{Harry Brignull}.}
  \bibinfo{year}{2013}\natexlab{}.
\newblock \showarticletitle{Dark Patterns: inside the interfaces designed to
  trick you}.
\newblock \bibinfo{journal}{\emph{The Verge}} (\bibinfo{year}{2013}).
\newblock


\bibitem[\protect\citeauthoryear{Brignull, Miquel, Rosenberg, and
  Offer}{Brignull et~al\mbox{.}}{2015}]%
        {Brignull2015-qs}
\bibfield{author}{\bibinfo{person}{Harry Brignull}, \bibinfo{person}{Marc
  Miquel}, \bibinfo{person}{Jeremy Rosenberg}, {and} \bibinfo{person}{James
  Offer}.} \bibinfo{year}{2015}\natexlab{}.
\newblock \bibinfo{title}{Dark {Patterns-User} Interfaces Designed to Trick
  People}.
\newblock
\newblock


\bibitem[\protect\citeauthoryear{Buwert}{Buwert}{2018}]%
        {Buwert2018-uw}
\bibfield{author}{\bibinfo{person}{Peter Buwert}.}
  \bibinfo{year}{2018}\natexlab{}.
\newblock \showarticletitle{Examining the Professional Codes of Design
  Organisations}. In \bibinfo{booktitle}{\emph{Proceedings of the Design
  Research Society}}.
\newblock
\urldef\tempurl%
\url{https://doi.org/10.21606/dma.2017.493}
\showDOI{\tempurl}


\bibitem[\protect\citeauthoryear{Carspecken}{Carspecken}{1996}]%
        {Carspecken1996-bq}
\bibfield{author}{\bibinfo{person}{P~F Carspecken}.}
  \bibinfo{year}{1996}\natexlab{}.
\newblock \bibinfo{booktitle}{\emph{Critical ethnography in educational
  research: A theoretical and practical guide}}.
\newblock \bibinfo{publisher}{Routledge}, \bibinfo{address}{New York}.
\newblock
\showISBNx{9780415904933}


\bibitem[\protect\citeauthoryear{Chromik, Eiband, V{\"o}lkel, and
  Buschek}{Chromik et~al\mbox{.}}{2019}]%
        {Chromik2019-rm}
\bibfield{author}{\bibinfo{person}{Michael Chromik}, \bibinfo{person}{Malin
  Eiband}, \bibinfo{person}{Sarah~Theres V{\"o}lkel}, {and}
  \bibinfo{person}{Daniel Buschek}.} \bibinfo{year}{2019}\natexlab{}.
\newblock \showarticletitle{Dark Patterns of Explainability, Transparency, and
  User Control for Intelligent Systems}. In \bibinfo{booktitle}{\emph{{IUI}
  Workshops}}. \bibinfo{publisher}{medien.ifi.lmu.de}.
\newblock


\bibitem[\protect\citeauthoryear{Coldewey}{Coldewey}{2018}]%
        {Coldewey2018-dx}
\bibfield{author}{\bibinfo{person}{Devin Coldewey}.}
  \bibinfo{year}{2018}\natexlab{}.
\newblock \showarticletitle{Students confront the unethical side of tech in
  `Designing for Evil' course}.
\newblock \bibinfo{journal}{\emph{TechCrunch}} (\bibinfo{date}{May}
  \bibinfo{year}{2018}).
\newblock
\urldef\tempurl%
\url{http://techcrunch.com/2018/05/29/students-confront-the-unethical-side-of-tech-in-designing-for-evil-course/}
\showURL{%
\tempurl}


\bibitem[\protect\citeauthoryear{Cowen and Keltner}{Cowen and Keltner}{2017}]%
        {Cowen2017-fn}
\bibfield{author}{\bibinfo{person}{Alan~S Cowen} {and} \bibinfo{person}{Dacher
  Keltner}.} \bibinfo{year}{2017}\natexlab{}.
\newblock \showarticletitle{Self-report captures 27 distinct categories of
  emotion bridged by continuous gradients}.
\newblock \bibinfo{journal}{\emph{Proc. Natl. Acad. Sci. U. S. A.}}
  \bibinfo{volume}{114}, \bibinfo{number}{38} (\bibinfo{date}{Sept.}
  \bibinfo{year}{2017}), \bibinfo{pages}{E7900--E7909}.
\newblock
\showISSN{0027-8424, 1091-6490}
\urldef\tempurl%
\url{https://doi.org/10.1073/pnas.1702247114}
\showDOI{\tempurl}


\bibitem[\protect\citeauthoryear{Deterding, Stenros, and Montola}{Deterding
  et~al\mbox{.}}{2020}]%
        {Deterding2020-fx}
\bibfield{author}{\bibinfo{person}{Christoph~Sebastian Deterding},
  \bibinfo{person}{Jaakko Stenros}, {and} \bibinfo{person}{Markus Montola}.}
  \bibinfo{year}{2020}\natexlab{}.
\newblock \showarticletitle{Against`` Dark Game Design Patterns''}. In
  \bibinfo{booktitle}{\emph{{DiGRA'20-Abstract} Proceedings of the 2020 {DiGRA}
  International Conference}}. \bibinfo{publisher}{eprints.whiterose.ac.uk}.
\newblock


\bibitem[\protect\citeauthoryear{Di~Geronimo, Braz, Fregnan, Palomba, and
  Bacchelli}{Di~Geronimo et~al\mbox{.}}{2020}]%
        {Di_Geronimo2020-mh}
\bibfield{author}{\bibinfo{person}{Linda Di~Geronimo}, \bibinfo{person}{Larissa
  Braz}, \bibinfo{person}{Enrico Fregnan}, \bibinfo{person}{Fabio Palomba},
  {and} \bibinfo{person}{Alberto Bacchelli}.} \bibinfo{year}{2020}\natexlab{}.
\newblock \showarticletitle{{UI} Dark Patterns and Where to Find Them: A Study
  on Mobile Applications and User Perception}. In
  \bibinfo{booktitle}{\emph{Proceedings of the 2020 {CHI} Conference on Human
  Factors in Computing Systems}} (Honolulu, HI, USA)
  \emph{(\bibinfo{series}{CHI '20})}. \bibinfo{publisher}{Association for
  Computing Machinery}, \bibinfo{address}{New York, NY, USA},
  \bibinfo{pages}{1--14}.
\newblock
\showISBNx{9781450367080}
\urldef\tempurl%
\url{https://doi.org/10.1145/3313831.3376600}
\showDOI{\tempurl}


\bibitem[\protect\citeauthoryear{Fogg}{Fogg}{2003}]%
        {Fogg2003-rg}
\bibfield{author}{\bibinfo{person}{B~J Fogg}.} \bibinfo{year}{2003}\natexlab{}.
\newblock \bibinfo{booktitle}{\emph{Persuasive Technology: Using Computers to
  Change What We Think and Do}}.
\newblock 1--282 pages.
\newblock
\showISBNx{9781558606432}
\showISSN{1569-1101}
\urldef\tempurl%
\url{https://doi.org/10.1016/B978-1-55860-643-2.X5000-8}
\showDOI{\tempurl}


\bibitem[\protect\citeauthoryear{Fritsch}{Fritsch}{2017}]%
        {Fritsch2017-re}
\bibfield{author}{\bibinfo{person}{Lothar Fritsch}.}
  \bibinfo{year}{2017}\natexlab{}.
\newblock \showarticletitle{Privacy dark patterns in identity management}. In
  \bibinfo{booktitle}{\emph{Open Identity Summit ({OID)}, 5-6 october 2017,
  Karlstad, Sweden.}} \bibinfo{publisher}{Gesellschaft f{\"u}r Informatik},
  \bibinfo{pages}{93--104}.
\newblock


\bibitem[\protect\citeauthoryear{Garland}{Garland}{1964}]%
        {Garland1964-cv}
\bibfield{author}{\bibinfo{person}{Ken Garland}.}
  \bibinfo{year}{1964}\natexlab{}.
\newblock \bibinfo{title}{First things first manifesto}.
\newblock
  \bibinfo{howpublished}{\url{http://www.designishistory.com/1960/first-things-first/}}.
\newblock
\urldef\tempurl%
\url{http://www.designishistory.com/1960/first-things-first/}
\showURL{%
\tempurl}


\bibitem[\protect\citeauthoryear{Gotterbarn, Brinkman, Flick, Kirkpatrick,
  Miller, Vazansky, and Wolf}{Gotterbarn et~al\mbox{.}}{2018}]%
        {Gotterbarn2018-ja}
\bibfield{author}{\bibinfo{person}{D~W Gotterbarn}, \bibinfo{person}{Bo
  Brinkman}, \bibinfo{person}{Catherine Flick}, \bibinfo{person}{Michael~S
  Kirkpatrick}, \bibinfo{person}{Keith Miller}, \bibinfo{person}{Kate
  Vazansky}, {and} \bibinfo{person}{Marty~J Wolf}.}
  \bibinfo{year}{2018}\natexlab{}.
\newblock \showarticletitle{{ACM} code of ethics and professional conduct}.
\newblock  (\bibinfo{year}{2018}).
\newblock
\urldef\tempurl%
\url{https://dora.dmu.ac.uk/handle/2086/16422}
\showURL{%
\tempurl}


\bibitem[\protect\citeauthoryear{Grassl, Schraffenberger, Zuiderveen~Borgesius,
  and Buijzen}{Grassl et~al\mbox{.}}{2020}]%
        {Grassl2020-lh}
\bibfield{author}{\bibinfo{person}{Paul Grassl}, \bibinfo{person}{Hanna
  Schraffenberger}, \bibinfo{person}{Frederik Zuiderveen~Borgesius}, {and}
  \bibinfo{person}{Moniek Buijzen}.} \bibinfo{year}{2020}\natexlab{}.
\newblock \bibinfo{title}{Dark and bright patterns in cookie consent requests}.
   (\bibinfo{date}{July} \bibinfo{year}{2020}).
\newblock
\urldef\tempurl%
\url{https://doi.org/10.31234/osf.io/gqs5h}
\showDOI{\tempurl}


\bibitem[\protect\citeauthoryear{Gray and Chivukula}{Gray and
  Chivukula}{2019}]%
        {Gray2019-ep}
\bibfield{author}{\bibinfo{person}{Colin~M Gray} {and}
  \bibinfo{person}{Shruthi~Sai Chivukula}.} \bibinfo{year}{2019}\natexlab{}.
\newblock \showarticletitle{Ethical Mediation in {UX} Practice}. In
  \bibinfo{booktitle}{\emph{Proceedings of the 2019 {CHI} Conference on Human
  Factors in Computing Systems - {CHI} '19}}. \bibinfo{publisher}{ACM Press}.
\newblock
\urldef\tempurl%
\url{https://doi.org/10.1145/3290605.3300408}
\showDOI{\tempurl}


\bibitem[\protect\citeauthoryear{Gray, Chivukula, and Lee}{Gray
  et~al\mbox{.}}{2020a}]%
        {Gray2020-zq}
\bibfield{author}{\bibinfo{person}{Colin~M Gray}, \bibinfo{person}{Shruthi~Sai
  Chivukula}, {and} \bibinfo{person}{Ahreum Lee}.}
  \bibinfo{year}{2020}\natexlab{a}.
\newblock \showarticletitle{What Kind of Work Do ``Asshole Designers'' Create?
  Describing Properties of Ethical Concern on Reddit}. In
  \bibinfo{booktitle}{\emph{Proceedings of the 2020 {ACM} Designing Interactive
  Systems Conference}} (Eindhoven, Netherlands) \emph{(\bibinfo{series}{DIS
  '20})}. \bibinfo{publisher}{Association for Computing Machinery},
  \bibinfo{address}{New York, NY, USA}, \bibinfo{pages}{61--73}.
\newblock
\showISBNx{9781450369749}
\urldef\tempurl%
\url{https://doi.org/10.1145/3357236.3395486}
\showDOI{\tempurl}


\bibitem[\protect\citeauthoryear{Gray, Kou, Battles, Hoggatt, and Toombs}{Gray
  et~al\mbox{.}}{2018}]%
        {Gray2018-or}
\bibfield{author}{\bibinfo{person}{Colin~M Gray}, \bibinfo{person}{Yubo Kou},
  \bibinfo{person}{Bryan Battles}, \bibinfo{person}{Joseph Hoggatt}, {and}
  \bibinfo{person}{Austin~L Toombs}.} \bibinfo{year}{2018}\natexlab{}.
\newblock \showarticletitle{{The Dark (Patterns) Side of {UX} Design}}. In
  \bibinfo{booktitle}{\emph{Proceedings of the 2018 {CHI} Conference on Human
  Factors in Computing Systems}} (Montreal QC, Canada)
  \emph{(\bibinfo{series}{CHI '18})}. \bibinfo{publisher}{dl.acm.org},
  \bibinfo{address}{New York, NY, USA}, \bibinfo{pages}{534:1--534:14}.
\newblock
\showISBNx{9781450356206}
\urldef\tempurl%
\url{https://doi.org/10.1145/3173574.3174108}
\showDOI{\tempurl}


\bibitem[\protect\citeauthoryear{Gray, Santos, Bielova, Toth, and
  Clifford}{Gray et~al\mbox{.}}{2020b}]%
        {Gray2020-zf}
\bibfield{author}{\bibinfo{person}{Colin~M Gray}, \bibinfo{person}{Cristiana
  Santos}, \bibinfo{person}{Nataliia Bielova}, \bibinfo{person}{Michael Toth},
  {and} \bibinfo{person}{Damian Clifford}.} \bibinfo{year}{2020}\natexlab{b}.
\newblock \bibinfo{title}{Dark Patterns and the Legal Requirements of Consent
  Banners: An Interaction Criticism Perspective}.
\newblock
\newblock
\showeprint[arxiv]{2009.10194}~[cs.HC]
\urldef\tempurl%
\url{http://arxiv.org/abs/2009.10194}
\showURL{%
\tempurl}


\bibitem[\protect\citeauthoryear{Greenberg, Boring, Vermeulen, and
  Dostal}{Greenberg et~al\mbox{.}}{2014}]%
        {Greenberg2014-dg}
\bibfield{author}{\bibinfo{person}{Saul Greenberg}, \bibinfo{person}{Sebastian
  Boring}, \bibinfo{person}{Jo Vermeulen}, {and} \bibinfo{person}{Jakub
  Dostal}.} \bibinfo{year}{2014}\natexlab{}.
\newblock \showarticletitle{Dark Patterns in Proxemic Interactions: A Critical
  Perspective}. In \bibinfo{booktitle}{\emph{Proceedings of the 2014 Conference
  on Designing Interactive Systems}} (Vancouver, BC, Canada)
  \emph{(\bibinfo{series}{DIS '14})}. \bibinfo{publisher}{ACM},
  \bibinfo{address}{New York, NY, USA}, \bibinfo{pages}{523--532}.
\newblock
\showISBNx{9781450329026}
\urldef\tempurl%
\url{https://doi.org/10.1145/2598510.2598541}
\showDOI{\tempurl}


\bibitem[\protect\citeauthoryear{Kelly}{Kelly}{2019}]%
        {Kelly2019-me}
\bibfield{author}{\bibinfo{person}{Makena Kelly}.}
  \bibinfo{year}{2019}\natexlab{}.
\newblock \bibinfo{title}{Big Tech's `dark patterns' could be outlawed under
  new Senate bill}.
\newblock
  \bibinfo{howpublished}{\url{https://www.theverge.com/2019/4/9/18302199/big-tech-dark-patterns-senate-bill-detour-act-facebook-google-amazon-twitter}}.
\newblock
\urldef\tempurl%
\url{https://www.theverge.com/2019/4/9/18302199/big-tech-dark-patterns-senate-bill-detour-act-facebook-google-amazon-twitter}
\showURL{%
\tempurl}
\newblock
\shownote{Accessed: 2020-9-19.}


\bibitem[\protect\citeauthoryear{Lacey and Caudwell}{Lacey and
  Caudwell}{2019}]%
        {Lacey2019-cy}
\bibfield{author}{\bibinfo{person}{C Lacey} {and} \bibinfo{person}{C
  Caudwell}.} \bibinfo{year}{2019}\natexlab{}.
\newblock \showarticletitle{Cuteness as a `Dark Pattern' in Home Robots}. In
  \bibinfo{booktitle}{\emph{2019 14th {ACM/IEEE} International Conference on
  {Human-Robot} Interaction ({HRI})}}. \bibinfo{pages}{374--381}.
\newblock
\showISSN{2167-2148}
\urldef\tempurl%
\url{https://doi.org/10.1109/HRI.2019.8673274}
\showDOI{\tempurl}


\bibitem[\protect\citeauthoryear{Luguri and Strahilevitz}{Luguri and
  Strahilevitz}{2019}]%
        {Luguri2019-bg}
\bibfield{author}{\bibinfo{person}{Jamie Luguri} {and} \bibinfo{person}{Lior
  Strahilevitz}.} \bibinfo{year}{2019}\natexlab{}.
\newblock \bibinfo{title}{Shining a Light on Dark Patterns}.
  (\bibinfo{date}{Aug.} \bibinfo{year}{2019}).
\newblock
\urldef\tempurl%
\url{https://doi.org/10.2139/ssrn.3431205}
\showDOI{\tempurl}


\bibitem[\protect\citeauthoryear{MacDonald}{MacDonald}{2019}]%
        {MacDonald2019-nq}
\bibfield{author}{\bibinfo{person}{Diana MacDonald}.}
  \bibinfo{year}{2019}\natexlab{}.
\newblock \showarticletitle{Anti-patterns and dark patterns}.
\newblock In \bibinfo{booktitle}{\emph{Practical {UI} Patterns for Design
  Systems: {Fast-Track} Interaction Design for a Seamless User Experience}},
  \bibfield{editor}{\bibinfo{person}{Diana MacDonald}} (Ed.).
  \bibinfo{publisher}{Apress}, \bibinfo{address}{Berkeley, CA},
  \bibinfo{pages}{193--221}.
\newblock
\showISBNx{9781484249383}
\urldef\tempurl%
\url{https://doi.org/10.1007/978-1-4842-4938-3\_5}
\showDOI{\tempurl}


\bibitem[\protect\citeauthoryear{Maier and Harr}{Maier and Harr}{2020}]%
        {Maier2020-yk}
\bibfield{author}{\bibinfo{person}{Maximilian Maier} {and}
  \bibinfo{person}{Rikard Harr}.} \bibinfo{year}{2020}\natexlab{}.
\newblock \showarticletitle{Dark Design Patterns: An {End-User} Perspective}.
\newblock \bibinfo{journal}{\emph{Human Technology}} \bibinfo{volume}{16},
  \bibinfo{number}{2} (\bibinfo{year}{2020}), \bibinfo{pages}{170--199}.
\newblock
\urldef\tempurl%
\url{https://doi.org/10.17011/ht/urn.202008245641}
\showDOI{\tempurl}


\bibitem[\protect\citeauthoryear{Mathur, Acar, Friedman, Lucherini, Mayer,
  Chetty, and Narayanan}{Mathur et~al\mbox{.}}{2019}]%
        {Mathur2019-ea}
\bibfield{author}{\bibinfo{person}{Arunesh Mathur}, \bibinfo{person}{Gunes
  Acar}, \bibinfo{person}{Michael~J Friedman}, \bibinfo{person}{Elena
  Lucherini}, \bibinfo{person}{Jonathan Mayer}, \bibinfo{person}{Marshini
  Chetty}, {and} \bibinfo{person}{Arvind Narayanan}.}
  \bibinfo{year}{2019}\natexlab{}.
\newblock \showarticletitle{Dark Patterns at Scale: Findings from a Crawl of
  {11K} Shopping Websites}.
\newblock  (\bibinfo{date}{July} \bibinfo{year}{2019}).
\newblock
\showeprint[arxiv]{1907.07032}~[cs.HC]


\bibitem[\protect\citeauthoryear{Mirnig and Tscheligi}{Mirnig and
  Tscheligi}{2017}]%
        {Mirnig2017-ij}
\bibfield{author}{\bibinfo{person}{Alexander Mirnig} {and}
  \bibinfo{person}{Manfred Tscheligi}.} \bibinfo{year}{2017}\natexlab{}.
\newblock \showarticletitle{(Don't) Join the Dark Side: An Initial Analysis and
  Classification of Regular, Anti-, and Dark Patterns}. In
  \bibinfo{booktitle}{\emph{{PATTERNS} 2017: Proceedings of the 9th
  International Conference on Pervasive Patterns and Applications}}.
  \bibinfo{pages}{65--71}.
\newblock


\bibitem[\protect\citeauthoryear{Moser}{Moser}{2020}]%
        {Moser2020-sf}
\bibfield{author}{\bibinfo{person}{Carol Moser}.}
  \bibinfo{year}{2020}\natexlab{}.
\newblock \emph{\bibinfo{title}{Impulse Buying: Designing for {Self-Control}
  with E-commerce}}.
\newblock \bibinfo{thesistype}{Ph.D. Dissertation}. \bibinfo{school}{University
  of Michigal}.
\newblock


\bibitem[\protect\citeauthoryear{Narayanan, Mathur, Chetty, and
  Kshirsagar}{Narayanan et~al\mbox{.}}{2020}]%
        {Narayanan2020-va}
\bibfield{author}{\bibinfo{person}{Arvind Narayanan}, \bibinfo{person}{Arunesh
  Mathur}, \bibinfo{person}{Marshini Chetty}, {and} \bibinfo{person}{Mihir
  Kshirsagar}.} \bibinfo{year}{2020}\natexlab{}.
\newblock \showarticletitle{Dark Patterns: Past, Present, and Future}.
\newblock \bibinfo{journal}{\emph{Queueing Systems. Theory and Applications}}
  \bibinfo{volume}{18}, \bibinfo{number}{2} (\bibinfo{date}{April}
  \bibinfo{year}{2020}), \bibinfo{pages}{67--92}.
\newblock
\showISSN{0257-0130, 1542-7730}
\urldef\tempurl%
\url{https://doi.org/10.1145/3400899.3400901}
\showDOI{\tempurl}


\bibitem[\protect\citeauthoryear{Nodder}{Nodder}{2013}]%
        {Nodder2013-rw}
\bibfield{author}{\bibinfo{person}{Chris Nodder}.}
  \bibinfo{year}{2013}\natexlab{}.
\newblock \bibinfo{booktitle}{\emph{Evil by Design: Interaction Design to Lead
  Us into Temptation}}.
\newblock \bibinfo{publisher}{John Wiley \& Sons, Inc.},
  \bibinfo{address}{Indianapolis, IN}. 303 pages.
\newblock
\showISBNx{9781118654811}
\urldef\tempurl%
\url{https://market.android.com/details?id=book-46Wl1G9yJUoC}
\showURL{%
\tempurl}


\bibitem[\protect\citeauthoryear{Nouwens, Liccardi, Veale, Karger, and
  Kagal}{Nouwens et~al\mbox{.}}{2020}]%
        {Nouwens2020-ij}
\bibfield{author}{\bibinfo{person}{Midas Nouwens}, \bibinfo{person}{Ilaria
  Liccardi}, \bibinfo{person}{Michael Veale}, \bibinfo{person}{David Karger},
  {and} \bibinfo{person}{Lalana Kagal}.} \bibinfo{year}{2020}\natexlab{}.
\newblock \showarticletitle{Dark Patterns after the {GDPR}: Scraping Consent
  Pop-ups and Demonstrating their Influence}. In
  \bibinfo{booktitle}{\emph{Proceedings of the 2020 {CHI} Conference on Human
  Factors in Computing Systems}} (Honolulu, HI, USA)
  \emph{(\bibinfo{series}{CHI '20})}. \bibinfo{publisher}{Association for
  Computing Machinery}, \bibinfo{address}{New York, NY, USA},
  \bibinfo{pages}{1--13}.
\newblock
\showISBNx{9781450367080}
\urldef\tempurl%
\url{https://doi.org/10.1145/3313831.3376321}
\showDOI{\tempurl}


\bibitem[\protect\citeauthoryear{Read}{Read}{2018}]%
        {Read2018-cr}
\bibfield{author}{\bibinfo{person}{Brandon Read}.}
  \bibinfo{year}{2018}\natexlab{}.
\newblock \bibinfo{title}{{TurboTax}: a critical analysis \& {UX} design
  teardown - {UX} Collective}.
\newblock
  \bibinfo{howpublished}{\url{https://uxdesign.cc/turbotax-design-1a37356adc61}}.
\newblock
\urldef\tempurl%
\url{https://uxdesign.cc/turbotax-design-1a37356adc61}
\showURL{%
\tempurl}
\newblock
\shownote{Accessed: 2020-10-10.}


\bibitem[\protect\citeauthoryear{Shilton}{Shilton}{2013}]%
        {Shilton2013-dq}
\bibfield{author}{\bibinfo{person}{Katie Shilton}.}
  \bibinfo{year}{2013}\natexlab{}.
\newblock \showarticletitle{{Values Levers: Building Ethics into Design}}.
\newblock \bibinfo{journal}{\emph{Science, technology \& human values}}
  \bibinfo{volume}{38}, \bibinfo{number}{3} (\bibinfo{date}{May}
  \bibinfo{year}{2013}), \bibinfo{pages}{374--397}.
\newblock
\showISSN{0162-2439}
\urldef\tempurl%
\url{https://doi.org/10.1177/0162243912436985}
\showDOI{\tempurl}


\bibitem[\protect\citeauthoryear{Shilton}{Shilton}{2018}]%
        {Shilton2018-sw}
\bibfield{author}{\bibinfo{person}{Katie Shilton}.}
  \bibinfo{year}{2018}\natexlab{}.
\newblock \showarticletitle{Engaging Values Despite Neutrality: Challenges and
  Approaches to Values Reflection during the Design of Internet
  Infrastructure}.
\newblock \bibinfo{journal}{\emph{Science, technology \& human values}}
  \bibinfo{volume}{43}, \bibinfo{number}{2} (\bibinfo{date}{March}
  \bibinfo{year}{2018}), \bibinfo{pages}{016224391771486}.
\newblock
\showISSN{0162-2439}
\urldef\tempurl%
\url{https://doi.org/10.1177/0162243917714869}
\showDOI{\tempurl}


\bibitem[\protect\citeauthoryear{Soe, Nordberg, Guribye, and Slavkovik}{Soe
  et~al\mbox{.}}{2020}]%
        {Soe2020-se}
\bibfield{author}{\bibinfo{person}{Than~Htut Soe}, \bibinfo{person}{Oda~Elise
  Nordberg}, \bibinfo{person}{Frode Guribye}, {and} \bibinfo{person}{Marija
  Slavkovik}.} \bibinfo{year}{2020}\natexlab{}.
\newblock \showarticletitle{Circumvention by design---dark patterns in cookie
  consents for online news outlets}.
\newblock  (\bibinfo{date}{June} \bibinfo{year}{2020}).
\newblock
\showeprint[arxiv]{2006.13985}~[cs.HC]
\urldef\tempurl%
\url{http://arxiv.org/abs/2006.13985}
\showURL{%
\tempurl}


\bibitem[\protect\citeauthoryear{Steen}{Steen}{2015}]%
        {Steen2015-qw}
\bibfield{author}{\bibinfo{person}{Marc Steen}.}
  \bibinfo{year}{2015}\natexlab{}.
\newblock \showarticletitle{Upon Opening the Black Box and Finding It Full:
  Exploring the Ethics in Design Practices}.
\newblock \bibinfo{journal}{\emph{Science, technology \& human values}}
  \bibinfo{volume}{40}, \bibinfo{number}{3} (\bibinfo{date}{May}
  \bibinfo{year}{2015}), \bibinfo{pages}{389--420}.
\newblock
\showISSN{0162-2439}
\urldef\tempurl%
\url{https://doi.org/10.1177/0162243914547645}
\showDOI{\tempurl}


\bibitem[\protect\citeauthoryear{Susser, Roessler, and Nissenbaum}{Susser
  et~al\mbox{.}}{2019}]%
        {Susser2019-dm}
\bibfield{author}{\bibinfo{person}{Daniel Susser}, \bibinfo{person}{Beate
  Roessler}, {and} \bibinfo{person}{Helen Nissenbaum}.}
  \bibinfo{year}{2019}\natexlab{}.
\newblock \showarticletitle{Technology, autonomy, and manipulation}.
\newblock \bibinfo{journal}{\emph{Internet Policy Review}} \bibinfo{volume}{8},
  \bibinfo{number}{2} (\bibinfo{year}{2019}).
\newblock
\urldef\tempurl%
\url{https://doi.org/10.14763/2019.2.1410}
\showDOI{\tempurl}


\bibitem[\protect\citeauthoryear{Tollady}{Tollady}{2016}]%
        {Tollady2016-pa}
\bibfield{author}{\bibinfo{person}{Benjamin Tollady}.}
  \bibinfo{year}{2016}\natexlab{}.
\newblock \bibinfo{title}{Under the influence: Dark patterns and the power of
  persuasive design}.
\newblock
  \bibinfo{howpublished}{\url{https://uxmastery.com/dark-patterns-and-the-power-of-persuasive-design/}}.
\newblock


\bibitem[\protect\citeauthoryear{Van~Dijk}{Van~Dijk}{2006}]%
        {van2006discourse}
\bibfield{author}{\bibinfo{person}{Teun~A Van~Dijk}.}
  \bibinfo{year}{2006}\natexlab{}.
\newblock \showarticletitle{Discourse and manipulation}.
\newblock \bibinfo{journal}{\emph{Discourse \& society}} \bibinfo{volume}{17},
  \bibinfo{number}{3} (\bibinfo{year}{2006}), \bibinfo{pages}{359--383}.
\newblock


\bibitem[\protect\citeauthoryear{Vance}{Vance}{2016}]%
        {Vance2016-yv}
\bibfield{author}{\bibinfo{person}{Alyssa Vance}.}
  \bibinfo{year}{2016}\natexlab{}.
\newblock \bibinfo{title}{Dark Patterns by the Boston Globe}.
\newblock
  \bibinfo{howpublished}{\url{https://rationalconspiracy.com/2016/04/24/dark-patterns-by-the-boston-globe/}}.
\newblock


\bibitem[\protect\citeauthoryear{Waldman}{Waldman}{2020}]%
        {Waldman2020-km}
\bibfield{author}{\bibinfo{person}{Ari~Ezra Waldman}.}
  \bibinfo{year}{2020}\natexlab{}.
\newblock \showarticletitle{Cognitive biases, dark patterns, and the 'privacy
  paradox'}.
\newblock \bibinfo{journal}{\emph{Current opinion in psychology}}
  \bibinfo{volume}{31} (\bibinfo{date}{Feb.} \bibinfo{year}{2020}),
  \bibinfo{pages}{105--109}.
\newblock
\showISSN{2352-2518, 2352-250X}
\urldef\tempurl%
\url{https://doi.org/10.1016/j.copsyc.2019.08.025}
\showDOI{\tempurl}


\bibitem[\protect\citeauthoryear{Watson, Clark, and Tellegen}{Watson
  et~al\mbox{.}}{1988}]%
        {Watson1988-po}
\bibfield{author}{\bibinfo{person}{David Watson}, \bibinfo{person}{Lee~Anna
  Clark}, {and} \bibinfo{person}{Auke Tellegen}.}
  \bibinfo{year}{1988}\natexlab{}.
\newblock \showarticletitle{Development and validation of brief measures of
  positive and negative affect: The {PANAS} scales}.
\newblock \bibinfo{journal}{\emph{Journal of personality and social
  psychology}} \bibinfo{volume}{54}, \bibinfo{number}{6}
  (\bibinfo{year}{1988}), \bibinfo{pages}{1063--1070}.
\newblock
\showISSN{0022-3514, 1939-1315}
\urldef\tempurl%
\url{https://doi.org/10.1037/0022-3514.54.6.1063}
\showDOI{\tempurl}


\bibitem[\protect\citeauthoryear{Weinmann, Schneider, and Brocke}{Weinmann
  et~al\mbox{.}}{2016}]%
        {Weinmann2016-oo}
\bibfield{author}{\bibinfo{person}{Markus Weinmann}, \bibinfo{person}{Christoph
  Schneider}, {and} \bibinfo{person}{Jan~Vom Brocke}.}
  \bibinfo{year}{2016}\natexlab{}.
\newblock \showarticletitle{Digital Nudging}.
\newblock \bibinfo{journal}{\emph{Business \& Information Systems Engineering}}
  \bibinfo{volume}{58}, \bibinfo{number}{6} (\bibinfo{date}{Dec.}
  \bibinfo{year}{2016}), \bibinfo{pages}{433--436}.
\newblock
\showISSN{1867-0202}
\urldef\tempurl%
\url{https://doi.org/10.1007/s12599-016-0453-1}
\showDOI{\tempurl}


\bibitem[\protect\citeauthoryear{Wilkinson}{Wilkinson}{2013}]%
        {wilkinson2013nudging}
\bibfield{author}{\bibinfo{person}{T~Martin Wilkinson}.}
  \bibinfo{year}{2013}\natexlab{}.
\newblock \showarticletitle{Nudging and manipulation}.
\newblock \bibinfo{journal}{\emph{Political Studies}} \bibinfo{volume}{61},
  \bibinfo{number}{2} (\bibinfo{year}{2013}), \bibinfo{pages}{341--355}.
\newblock


\bibitem[\protect\citeauthoryear{Wong}{Wong}{2020}]%
        {Wong2020-ja}
\bibfield{author}{\bibinfo{person}{Henry Wong}.}
  \bibinfo{year}{2020}\natexlab{}.
\newblock \bibinfo{title}{Inside the deceptive design world of dark patterns}.
\newblock
  \bibinfo{howpublished}{\url{https://www.designweek.co.uk/issues/9-15-march-2020/dark-patterns-design/}}.
\newblock
\newblock
\shownote{Accessed: 2020-3-10.}


\bibitem[\protect\citeauthoryear{Wong and Mulligan}{Wong and Mulligan}{2019}]%
        {Wong2019-rc}
\bibfield{author}{\bibinfo{person}{Richmond~Y Wong} {and}
  \bibinfo{person}{Deirdre~K Mulligan}.} \bibinfo{year}{2019}\natexlab{}.
\newblock \showarticletitle{Bringing Design to the Privacy Table: Broadening
  ``Design'' in ``Privacy by Design'' Through the Lens of {HCI}}. In
  \bibinfo{booktitle}{\emph{Proceedings of the 2019 {CHI} Conference on Human
  Factors in Computing Systems}} \emph{(\bibinfo{series}{CHI '19})}.
  \bibinfo{publisher}{ACM Press}, \bibinfo{address}{New York, NY}.
\newblock
\urldef\tempurl%
\url{https://doi.org/10.1145/3290605.3300492}
\showDOI{\tempurl}


\bibitem[\protect\citeauthoryear{Wong, Mulligan, Van~Wyk, Pierce, and
  Chuang}{Wong et~al\mbox{.}}{2017}]%
        {Wong2017-wt}
\bibfield{author}{\bibinfo{person}{Richmond~Y Wong}, \bibinfo{person}{Deirdre~K
  Mulligan}, \bibinfo{person}{Ellen Van~Wyk}, \bibinfo{person}{James Pierce},
  {and} \bibinfo{person}{John Chuang}.} \bibinfo{year}{2017}\natexlab{}.
\newblock \showarticletitle{Eliciting Values Reflections by Engaging Privacy
  Futures Using Design Workbooks}.
\newblock \bibinfo{journal}{\emph{Proceedings of the ACM on Human-Computer
  Interaction}} \bibinfo{volume}{1}, \bibinfo{number}{CSCW}
  (\bibinfo{date}{Dec.} \bibinfo{year}{2017}), \bibinfo{pages}{Article No.
  111}.
\newblock
\showISSN{2573-0142}
\urldef\tempurl%
\url{https://doi.org/10.1145/3134746}
\showDOI{\tempurl}


\bibitem[\protect\citeauthoryear{Zagal, Bj{\"o}rk, and Lewis}{Zagal
  et~al\mbox{.}}{2013}]%
        {Zagal2013-ms}
\bibfield{author}{\bibinfo{person}{Jos{\'e}~P Zagal}, \bibinfo{person}{Staffan
  Bj{\"o}rk}, {and} \bibinfo{person}{Chris Lewis}.}
  \bibinfo{year}{2013}\natexlab{}.
\newblock \showarticletitle{Dark Patterns in the Design of Games}. In
  \bibinfo{booktitle}{\emph{Foundations of Digital Games 2013}}.
  \bibinfo{publisher}{diva-portal.org}.
\newblock


\bibitem[\protect\citeauthoryear{Zuboff}{Zuboff}{2019}]%
        {Zuboff2019-kz}
\bibfield{author}{\bibinfo{person}{Shoshana Zuboff}.}
  \bibinfo{year}{2019}\natexlab{}.
\newblock \bibinfo{booktitle}{\emph{The Age of Surveillance Capitalism: The
  Fight for a Human Future at the New Frontier of Power}}.
  Vol.~\bibinfo{volume}{4}.
\newblock \bibinfo{publisher}{PublicAffairs}.
\newblock
\showISBNx{9781610395694}


\end{thebibliography}

\end{document}